\documentclass[twocolumn]{aastex63}

\newcommand{\sna}{SN~1987A}

\newcommand{\magab}{~mag$_\mathrm{AB}$}
\newcommand{\ctsps}{~counts~s$^{-1}$}
\newcommand{\kmps}{~km~s$^{-1}$}
\newcommand{\ergps}{~erg~s$^{-1}$}
\newcommand{\ergpspcm}{~erg~s$^{-1}$~cm$^{-2}$}

\newcommand{\xrt}{XT}

\newcommand{\Msun}{M$_{\sun}$}
\newcommand{\Rsun}{R$_{\sun}$}
\newcommand{\Lsun}{L$_{\sun}$}

\newcommand{\xray}{X\nobreakdash-ray}
\newcommand{\xmm}{\mbox{\textit{XMM-Newton}}}
\newcommand{\rosat}{\mbox{\textit{ROSAT}}}
\newcommand{\chandra}{\mbox{\textit{Chandra}}}
\newcommand{\swift}{\mbox{\textit{Swift}}}
\newcommand{\erosita}{\mbox{\textit{eROSITA}}}
\newcommand{\cat}{\mbox{3XMM-DR8}}

\newcommand{\rband}{\textit{r}\nobreakdash-band}

\newcommand{\zband}{\textit{z}\nobreakdash-band}
\newcommand{\jband}{\textit{J}\nobreakdash-band}

\usepackage{amsmath}
\makeatletter
\let\ftype@table\ftype@figure
\makeatother

\received{April 20, 2020}
\revised{May 7, 2020}
\accepted{May 7, 2020}
\submitjournal{ApJ}
\shorttitle{Blasts from the Past} 
\shortauthors{Alp \& Larsson}

\begin{document}

\title{Blasts from the Past:\\
  Supernova Shock Breakouts among X-Ray Transients in the
  \textit{XMM-Newton} Archive}

\author[0000-0002-0427-5592]{Dennis Alp}
\affiliation{Department of Physics, KTH Royal Institute of Technology,
  and The Oskar Klein Centre, SE\nobreakdash-10691 Stockholm, Sweden}             

\author[0000-0003-0065-2933]{Josefin Larsson}
\affiliation{Department of Physics, KTH Royal Institute of Technology,
  and The Oskar Klein Centre, SE\nobreakdash-10691 Stockholm, Sweden}             

\begin{abstract}
  The first electromagnetic signal from a supernova (SN) is released
  when the shock crosses the progenitor surface. This shock breakout
  (SBO) emission provides constraints on progenitor and explosion
  properties. Observationally, SBOs appear as minute to hour-long
  extragalactic \xray{} transients. They are challenging to detect and
  only one SBO has been observed to date. Here, we search the
  \textit{XMM-Newton} archive and find twelve new SN SBO
  candidates. We identify host galaxies to nine of these at estimated
  redshifts of 0.1--1. The SBO candidates have energies of
  ${\sim}10^{46}$~erg, timescales of 30--3000~s, and temperatures of
  0.1--1~keV. They are all consistent with being SN SBOs, but some may
  be misidentified Galactic foreground sources or other extragalactic
  objects. SBOs from blue supergiants agree well with most of the
  candidates. However, a few could be SBOs from Wolf--Rayet stars
  surrounded by dense circumstellar media, whereas two are more
  naturally explained as SBOs from red supergiants. The observations
  tentatively support non-spherical SBOs and are in agreement with
  asymmetries predicted by recent three-dimensional SN explosion
  simulations. \erosita{} may detect ${\sim}$2 SBOs per year, which
  could be detected in live analyses and promptly followed up.
\end{abstract}


\section{Introduction}\label{sec:intro}
The shock breakout (SBO) emission is the first electromagnetic signal
from a supernova \citep[SN;][]{waxman17b, levinson19}. This emission
carries information about the structure of the progenitor star. The
most important physical properties that can be constrained are the
progenitor radius, asymmetries, and final mass-loss history. This
means that SBOs offer a unique avenue to probe SN progenitors
\citep{smartt09} and the SN explosion mechanism \citep{janka16}.

The SBO emission is released when the radiation mediated shock from
the SN explosion crosses the surface of the star. However, if the
circumstellar medium (CSM) is sufficiently dense, the shock can
propagate into the CSM and result in a longer ``CSM breakout''. SBOs
from the progenitor surface typically peak in \xray{}s, and evolve
into ultraviolet and optical as the envelope cools and the ejecta
expand. Extended CSM breakouts and the post-breakout cooling phase
evolve on timescales of $>1$~day and are easier to detect. These
timescales are within the reach of recent wide-field optical SN
surveys (\citealt{waxman17b} and references therein). Henceforth, we
focus on \xray{} SBOs that evolve on timescales much shorter than
1~day. The initial SBO and post-breakout cooling phase precede the
commonly observed, months-long SN emission that is powered by
reprocessed radioactive decay.

The observable properties of SN SBOs are expected to be different for
different progenitor types. Typical SBO energies, timescales, and
temperatures \citep{matzner99, nakar10, sapir13} for red supergiants
(RSGs) are $10^{48}$~erg, 1000~s, and 0.03~keV, respectively. The
corresponding values are $10^{46.5}$~erg, 100~s, and 0.3~keV for blue
supergiants (BSGs); and $10^{45}$~erg, 10~s, and 3~keV for Wolf--Rayet
(WR) stars. This assumes spherical symmetry, as well as negligible
effects of the CSM. SBOs from RSGs are very soft and are expected to
be heavily absorbed by the interstellar medium (ISM). In contrast, WR
SBOs are much harder, but the total number of emitted photons is
several orders of magnitude lower. Both the increased temperature
(higher photon energies) and lower total energy reduce the total
number of emitted photons, which makes WR SBOs difficult to
detect. BSGs are the progenitor type that is most likely to be
detected due to the trade-off between peak energy and number of
emitted photons \citep{calzavara04, sapir13, waxman17b}. Finally, we
note that thermonuclear SNe are expected to produce SBOs with
temperatures of ${\sim}20$~keV. These SBOs are not expected to be
detectable because the timescales are approximately 10~ms and the
total energies are on the order of $10^{40}$~erg
\citep{piro10,nakar10}.

Theoretical aspects of SBOs have been studied for several
decades. Early works explored the emission from non-relativistic SBOs
with simplified physics and for specific progenitor structures (e.g.\
\citealt{weaver76, klein78, ensman92}). More recent works have focused
on developing models for more realistic progenitor structures with
more detailed treatments of the relevant physics \citep{katz10,
  nakar10, sapir11, katz12, sapir13, ito20}. Notably, several
theoretical works have stressed that many SBOs should have been
serendipitously observed by previous and existing \xray{} telescopes
\citep{klein78, calzavara04, sapir13, sapir14, waxman17b}. The
telescopes most likely to have detected SBOs are \xmm{}, which has
been in service since 2000, followed by \rosat{}, which collected data
between 1990 and 1999.

Despite the significant theoretical efforts, few observational
searches have been performed, and only the SBO from SN~2008D has been
detected \citep{soderberg08}. The SBO detection from SN~2008D was
serendipitously detected as a bright \xray{} transient in a scheduled
\swift{} observation of NGC~2770. In addition to the direct detection
of SN~2008D, the SBO from SN~1987A was indirectly observed by the
effects of the SBO on the CSM \citep{ensman92, blinnikov00}. The only
systematic searches in archival \xray{} data for transients with
SBO-like properties were of the \rosat{} archive \citep{vikhlinin98,
  greiner00}. Interestingly, \citet{vikhlinin98} reported a number of
candidates but was unable to securely identify their origins. While no
systematic searches for SBOs have been carried out recently,
transients on longer timescales have been studied, primarily in the
\chandra{} archive \citep{bauer17, xue19, yang19}.

In this paper, we search for SN SBOs in archival \xmm{}
data. Observationally, this implies that we search all public \xmm{}
observations for \xray{} transients on timescales shorter than
approximately 10~ks. To our knowledge, the \xmm{} data have not been
systematically searched for this kind of transient, even though
several SBOs are predicted to be present in the data. We aim to
identify SBO candidates and investigate the implications for the SN
progenitors.

This paper is organized as follows. We describe the SBO search and
identification process in Section~\ref{sec:indentification}, the host
galaxies in Section~\ref{sec:hosts}, and provide details related to
the data reduction of the final candidates in
Section~\ref{sec:obs}. The spectral fitting is described in
Section~\ref{sec:fit} and observed properties of the SBOs are
presented in Section~\ref{sec:observed}. We investigate the SBO
interpretation and its implications in Section~\ref{sec:discussion},
discuss contaminants and other potential astrophysical sources in
Section~\ref{sec:alt}, and conclude in Section~\ref{sec:conclusions}.

All uncertainties are 1$\sigma{}$ and one-sided limits are 3$\sigma{}$
unless stated otherwise. We adopt a flat $\Lambda{}$CDM cosmology with
a Hubble--Lema\^{i}tre constant $H_0 = 70$\kmps{}~Mpc$^{−1}$ and
$\Omega_\Lambda = 0.73$.

\section{Identifying SBO Candidates}\label{sec:indentification}
To find SN SBO candidates, we search for \xray{} transients with
typical timescales shorter than ${\sim}$10~ks in archival \xmm{}
data. These \xray{} transients can only be observed if they happen to
occur within the field of view (FoV) of an \xray{} telescope because
SBOs decay well before the optical signal from a SN is detected
(typically days or weeks later). \xmm{} \citep{jansen01} is the
telescope that is most likely to serendipitously observe SBOs because
of the combination of its high effective area, high angular
resolution, and large FoV of ${\sim}0.2$~deg$^2$. For theoretically
predicted BSG SBO properties (Section~\ref{sec:intro}), we expect our
search to be sensitive to a redshift of ${\sim}$1. This assumes
favorable observing conditions and a low level of absorption along the
line of sight.

Before performing a blind search for \xray{} transients, we crossmatch
all \xmm{} observations with known SNe and gamma-ray bursts (GRBs). In
the current context, we restrict the matches to \xray{} detections of
the SBO, long before the optical SN detection or simultaneous with the
prompt GRB phase. The purpose is to check if any known SNe or GRBs
have been serendipitously observed, but we find none. The catalog of
SNe is from the Latest Supernovae website\footnote{Previously known as
  ``Bright Supernovae'':
  \url{http://www.rochesterastronomy.org/snimages/sndateall.html}} and
the GRBs are from the \swift{}/BAT GRB Catalog \citep{lien16}.
The SN catalog covers the entire \xmm{} lifetime, while the GRB data
is limited to the lifetime of \swift{} because only GRBs with
sufficiently accurate positions can be crossmatched.

\subsection{Finding Transients}
To avoid detecting a large number of variable Galactic \xray{} sources
unrelated to SN SBOs, we immediately reject all sources that are close
to a star in Gaia or a known Galactic source in SIMBAD. More
precisely, we reject sources within 5~arcsec of an object in Gaia DR2
\citep{gaia18} with a (positive) parallax with a significance greater
than 3$\sigma$. Similarly for SIMBAD, we reject \xray{} sources if
there is a SIMBAD object within 5~arcsec classified as a star, white
dwarf, neutron star, black hole, or combinations thereof (i.e.\
binaries).

We develop a custom transient finder algorithm that we apply to all
observations, and also search for transients among the sources in the
\cat{} catalog \citep{rosen16}. The details of both search methods are
provided in Appendix~\ref{app:search}. The custom search algorithm is
the most general and detects any transients starting from the event
lists. The \cat{} search focuses on identifying transient behavior
among the cataloged \xray{} sources. The two algorithms are largely
redundant, but are complementary in some respects. The main difference
is that our custom search is performed on all public observations with
data archived at HEASARC as of 2019 November 11, whereas the latest
observation included in \cat{} is from 2017 November 30. We note that
the standard proprietary period is 1 year after data delivery, which
means that the availability of public data during the last 12--14
months of the time intervals is sparse.
Our custom algorithm is very simple and does not treat the background
and instrumental effects as carefully as \cat{}.

The two algorithms identify ${\sim}$11,000 transient sources and we
manually inspect the light curves, the time-integrated \xray{}
spectra, and images of these objects. The spectra and images are
primarily useful for recognizing instrumental artifacts, variability
caused by problematic extraction regions, and blended
sources. Instrumental artifacts are generally restricted to one of the
cameras and often affect individual pixels or columns. This is very
different from astrophysical sources, which are convolved by the point
spread function (PSF) of the telescope in all cameras. The screening
process up to this point reduces the number of \xray{} transients to
around 600.

\subsection{Source Classification}\label{sec:classification}
The final step is to separate the extragalactic transients from
Galactic foreground sources and artifacts. This separation essentially
finds all SBO candidates since SBOs are required to be associated with
galaxies. The remaining non-SBO transients are primarily flaring dwarf
stars that lack parallax in Gaia and are not identified in SIMBAD, or
instrumental artifacts that are difficult to identify in
time-integrated images.
To perform this final selection, we use time-resolved spectra and
images. Additionally, we search all catalogs available at VizieR
\citep{ochsenbein00} and inspect available sky images of the source
position (primarily \xray{}s, optical, and near-infrared; NIR). We
primarily use all HiPS sky images available through CDS/Aladin
\citep{bonnarel00, boch14}, but also check for data in MAST, ESO
Portal, and HEASARC in some cases. All objects that show clear signs
of being instrumental, Galactic, or persistent are discarded.

We separate stars from galaxies using the optical and NIR images. The
primary data used for host classification are from Pan-STARRS
\citep{chambers16}, DES \citep{abbott18}, and SkyMapper
\citep{wolf18}. We use morphological classifiers to identify extended
sources (galaxies)\footnote{e.g.\
  \url{https://outerspace.stsci.edu/display/PANSTARRS/How+to+separate+stars+and+galaxies}},
as well as stellaricity catalogs \citep{mcmahon13, tachibana18}.
However, since we are interested in accurate classifiers of a few
objects, we largely rely on visual inspection of available data for
each object. The manual classification of sources generally agrees
with automatic and objective classifiers, but allows us to identify
rare cases, such as likely misclassifications due to blending of
neighboring stars.

All objects that are possible SN SBOs are added to the final
sample. This decision is based on the light curve shape, duration,
spectral shape, and the probability that the source is extragalactic
based on available information. Furthermore, we require that the
initial rise and a substantial part of the decay is contained within
the observation. We discard sources that show persistent emission that
appears associated with the transient before the initial rise, which
would indicate that the source is a flaring persistent source. We do
not discard any transients because they are too short. This means that
the lower limit is effectively set by the temporal resolution of
\xmm{}, which is on the order of 100 ms for the most common observing
modes. We also do not discard any transients because they are too
long, but they are indirectly limited by the requirement of rising and
decaying within the exposure time (typically 10--130~ks). We do not
use quantitative requirements at this last stage because of the
difficulty of strictly defining the acceptable properties.

\begin{deluxetable*}{lrrrrccc}
  \caption{Identifiers of SBO Candidates\label{tab:identifiers}}
  \tablewidth{0pt}
  \tablehead{\colhead{\xrt{}} &  \colhead{$\alpha$} &  \colhead{$\delta$} &       \colhead{$l$} &       \colhead{$b$} &  \colhead{$\sigma$} & \colhead{Obs. ID} & \colhead{Time} \\
                   \colhead{} & \colhead{(\degr{})} & \colhead{(\degr{})} & \colhead{(\degr{})} & \colhead{(\degr{})} & \colhead{(\arcsec)} &        \colhead{} & \colhead{YYYY Mmm dd hh:mm:ss}} \startdata
                       161028 &         $263.23663$ &        $  43.51250$ &         $ 69.22769$ &         $ 32.05326$ &                 1.6 &        0781890401 & 2016 Oct 28 02:39:35 \\
                       151219 &         $173.53096$ &        $   0.87354$ &         $264.47274$ &         $ 57.85230$ &                 1.6 &        0770380401 & 2015 Dec 19 03:13:20 \\
                       110621 &         $ 37.89595$ &        $ -60.62934$ &         $283.01881$ &         $-52.45370$ &                 1.8 &        0675010401 & 2011 Jun 21 18:43:20 \\
                       030206 &         $ 29.28818$ &        $  37.62732$ &         $137.06247$ &         $-23.43464$ &                 1.3 &        0149780101 & 2003 Feb 06 03:52:30 \\
                       070618 &         $ 24.27520$ &        $ -12.95280$ &         $162.18014$ &         $-72.24921$ &                 1.4 &        0502020101 & 2007 Jun 18 21:03:55 \\
                       060207 &         $196.83165$ &        $ -40.46064$ &         $306.19786$ &         $ 22.31033$ &                 1.8 &        0300240501 & 2006 Feb 07 18:16:20 \\
                       100424 &         $321.79670$ &        $ -12.03906$ &         $ 40.02322$ &         $-39.95127$ &                 1.7 &        0604740101 & 2010 Apr 24 18:30:00 \\
                       151128 &         $167.07789$ &        $  -5.07517$ &         $261.31731$ &         $ 49.28100$ &                 2.0 &        0760380201 & 2015 Nov 28 23:20:00 \\\hline
                       050925 &         $311.43725$ &        $ -67.64702$ &         $327.21796$ &         $-35.70622$ &                 2.0 &        0300930301 & 2005 Sep 25 02:53:20 \\
                       160220 &         $204.19991$ &        $ -41.33736$ &         $312.01496$ &         $ 20.73297$ &                 1.6 &        0765041301 & 2016 Feb 20 07:28:20 \\
                       140811 &         $ 43.65356$ &        $  41.07430$ &         $146.61178$ &         $-16.06542$ &                 1.5 &        0743650701 & 2014 Aug 11 19:36:40 \\
                       040610 &         $169.53625$ &        $   7.70266$ &         $249.93237$ &         $ 60.57574$ &                 1.5 &        0203560201 & 2004 Jun 11 01:53:20 \\
  \enddata

  \tablecomments{Candidates are listed in descending order of
    confidence of being a SN SBO. The same ordering is used throughout
    the paper. The last four candidates may be misidentified Galactic
    foreground sources (Section~\ref{sec:dwarfs}). We provide
    celestial coordinates in both the equatorial ($\alpha$, $\delta$)
    and Galactic ($l$, $b$) systems. The parameter $\sigma{}$ refers
    to the astrometric uncertainty and the last column provides the
    start times of the transients in UTC.}
\end{deluxetable*}

We verify completeness of the search by inspecting the 100,000 most
variable light curves from \cat{} by eye to look for transients that
were potentially missed by the algorithms, and find one additional
faint candidate. The level of variability in this context is measured
by a likelihood test under the assumption that the light curve bin
uncertainties are Gaussian. Altogether, we find 12 SN SBO
candidates. Identifiers of the SN SBO candidates are provided in
Table~\ref{tab:identifiers}. We designate each candidate by ``\xrt{}''
followed by the date; the first two numbers correspond to the year,
the second two numbers to the month, and the last two numbers to the
day (analogous to GRBs). We choose to present the SBO candidates in
descending order of confidence (loosely defined) throughout this
paper.

\section{Host Galaxies}\label{sec:hosts}
\begin{deluxetable*}{lrrcccccccccccc}
  \caption{Host Candidates\label{tab:host_par}}
  \tablewidth{0pt}               
  \tablehead{\colhead{\xrt{}} &  \colhead{$\alpha$} &  \colhead{$\delta$} &    \colhead{Offset} & \colhead{$P_\mathrm{gal}$} & \colhead{$z$}          &               \colhead{$\mu$} & \colhead{Scale}               & \colhead{$M_r$}                & \colhead{Galaxy}  \\
                   \colhead{} & \colhead{(\degr{})} & \colhead{(\degr{})} & \colhead{(\arcsec)} & \colhead{}                 & \colhead{}             & \colhead{(mag$_\mathrm{AB}$)} & \colhead{(kpc~arcsec$^{-1}$)} & \colhead{(mag$_\mathrm{AB}$)}  & \colhead{}          }\startdata
  161028                      &         $263.23707$ &          $43.51231$ & 1.3                 & 1.00                       & $0.29$                 & 40.9                          & 4.4                           & $-20.2$                        & SB                  \\
  151219                      &         $173.53037$ &           $0.87409$ & 2.9                 & 0.99                       & $0.62$                 & 42.9                          & 6.9                           & $-21.4$                        & Sbc                 \\
  110621                      &          $37.89582$ &         $-60.62918$ & 0.6                 & 0.99                       & $0.095$                & 38.2                          & 1.8                           & $-18.6$                        & SB                  \\
  030206                      &          $29.28776$ &          $37.62768$ & 1.8                 & \nodata{}                  & $1.17$                 & 44.6                          & 8.4                           & $-22.5$                        & SB                  \\
  070618                      &           \nodata{} &           \nodata{} & \nodata{}           & \nodata{}                  & $0.37$                 & 41.5                          & 5.2                           & $>-17.4$                       & \nodata{}           \\
  060207                      &           \nodata{} &           \nodata{} & \nodata{}           & \nodata{}                  & $0.3$\tablenotemark{a} & 41.0                          & 4.5                           & $>-19.9$\tablenotemark{b}      & \nodata{}           \\
  100424                      &         $321.79659$ &         $-12.03900$ & 0.5                 & 1.00                       & $0.13$                 & 38.9                          & 2.3                           & $-18.6$                        & Sa                  \\
  151128                      &         $167.07885$ &          $-5.07495$ & 3.5                 & 1.00                       & $0.48$                 & 42.2                          & 6.0                           & $-21.9$                        & Sd                  \\\hline
  050925                      &         $311.43769$ &         $-67.64740$ & 1.5                 & 0.99                       & $0.3$\tablenotemark{a} & 41.0                          & 4.5                           & $-20.3$\tablenotemark{b}       & \nodata{}           \\
  160220                      &         $204.19926$ &         $-41.33718$ & 1.9                 & 0.90                       & $0.3$\tablenotemark{a} & 41.0                          & 4.5                           & $-20.7$\tablenotemark{b}       & \nodata{}           \\
  140811                      &          $43.65365$ &          $41.07406$ & 0.9                 & 0.75                       & $0.57$                 & 42.6                          & 6.6                           & $-22.0$                        & Sb                  \\
  040610                      &           \nodata{} &           \nodata{} & \nodata{}           & \nodata{}                  & $0.50$                 & 42.3                          & 6.2                           & $>-19.5$                       & \nodata{}           \\
  \enddata

  \tablecomments{Offset is the distance between the transient and the
    host galaxy. We note that the offsets are comparable to the
    uncertainties of the \xray{} positions
    (Table~\ref{tab:identifiers}). $P_\mathrm{gal}$ is the probability
    of the source being a galaxy based on morphology. These values are
    either from PS1\nobreakdash-PSC (Pan-STARRS data,
    \citealt{tachibana18}) or VHS (VISTA data,
    \citealt{mcmahon13}). For convenience, we also provide the
    distance expressed as the distance modulus ($\mu{}$), the angular
    scale, and the absolute magnitude ($M$; in \rband{} unless
    otherwise noted). The ``Galaxy'' column provides tentative galaxy
    classifiers based on the SED fitting; Sa, Sb, Sd, and Sbc are
    late-type spirals, whereas SB denotes starburst galaxies.}
  
  \tablenotetext{a}{Fiducial redshifts of 0.3 are used when
    insufficient information is available for a redshift estimate.}

  \tablenotetext{b}{These values are in the \jband{} because the
    \rband{} data is much shallower for these sources.}
\end{deluxetable*}

Table~\ref{tab:host_par} provides key parameters for the host galaxy
candidates associated with the 12 transients. For reference, we also
provide optical and NIR photometry of the hosts when available, and
limits otherwise, in Table~\ref{tab:host_mag}. Three of the transients
have no clearly identified host, and it is also possible that some of
the observed hosts are Galactic objects misclassified as galaxies. We
discuss potential misidentifications and alternative sources for the
X-ray transients in Section~\ref{sec:alt}. Throughout the rest of the
paper, we assume that the X-ray transients are SN SBOs.

The most important parameter for the inferred properties of the
transients is the redshift ($z$). Some of the redshifts are highly
uncertain and we stress that this uncertainty propagates into
parameters that depend on $z$. This mainly applies to the discussion
of inferred SN properties in Section~\ref{sec:sn_prop}. In contrast,
the observed properties in Section~\ref{sec:observed} are largely
independent of $z$, with the only effect being an energy shift of the
fitted temperature and host galaxy absorption. The magnitude of this
effect is relatively small ($\propto 1+z$) and does not qualitatively
alter the results.

To estimate the redshifts of the host galaxies, we primarily rely on
SED template fitting. Photometric redshift techniques are generally
more reliable if the Balmer (4000~\AA{}) or Lyman (912~\AA{}) break
falls between two broad filters. This means that the optical data
(Table~\ref{tab:host_mag}) is the most constraining for our
sources. Furthermore, we only use data from one survey (for each
object) to ensure that the photometry is homogeneous in the sense that
colors are correctly represented. For these reasons, we exclude the
NIR data from the SED fitting procedure. We are able to perform SED
fits to seven hosts, whereas the remaining five either lacked an
identified host or multi-band optical data.

In addition to the SED fitting, we also consider additional
independent information. The second most important constraint on $z$
comes from the redshifts of neighboring galaxies. We utilize this when
the SED fitting appears unreliable or if optical data of the host is
unavailable. After performing the SED fits and potentially including
data from neighbors, we check if the favored redshift implies a
reasonable absolute magnitude and radius for the galaxy \citep{shen03,
  blanton03, wolf03, ilbert05}. We also compare the distances with a
reconstruction of the large-scale structure (\citealt{lavaux19},
E.~Tsaprazi 2020, private communication). The reconstruction of the
large-scale structure in the local universe (out to $z={\sim}0.1$--1
depending on the line of sight) provides a measure of density as a
function of redshift along the line of sight. This is effectively the
likelihood of $z$ relative to the large-scale structure. The inclusion
of independent constraints means that we do not always choose the SED
fit with the lowest $\chi^2$ value, but rather analyze all data and
adopt the redshift that appears to be most consistent overall.
Finally, we note that the information from neighboring galaxies allows
us to estimate redshifts even if no host galaxy is directly
observed. This is only possible in some cases and is clearly highly
uncertain.

We use the \texttt{Le PHARE} code \citep{arnouts99, ilbert06} to
perform the SED fitting. 
The package includes a number of different galaxy SED libraries. We
try to fit all sources using the CWW-Kinney, AVEROIN, CFHTLS, and
COSMOS sets. Each set is fitted independently to explore the
systematic error introduced by choosing different SED libraries. This
also allows us to choose the SED that is most consistent with the
supplementary information. The CWW library \citep{coleman80} is based
on four observed spectra (Ell, Sbc, Scd, Irr) and is commonly used for
photometric redshift estimates. The CWW-Kinney sample extends the CWW
set with six starburst galaxies \citep{calzetti94, kinney96}. Both the
AVEROIN \citep{bruzual03, arnouts07} and CFHTLS \citep{ilbert06}
samples are essentially updated and refined versions of the CWW-Kinney
sample. Finally, the COSMOS sample \citep{ilbert09} is based on more
recent SEDs of three ellipticals and six spirals (S0, Sa, Sb, Sc, Sd,
Sdm, \citealt{polletta07}), as well as 12 starbursts
\citep{bruzual03}.

We conclude this section with brief notes on the redshift of
individual objects.
\begin{itemize}
\item Both \xrt{}~161028 ($z=0.29$) and 151219 ($z=0.62$) have
  photometric redshifts from SDSS-IV Data Release 16
  \citep{ahumada20}. We choose to adopt the SDSS values, which are
  very similar to the values obtained using our redshift estimation
  method ($0.28$ and $0.66$, respectively). This choice does not
  affect the results and facilitates comparisons with other works that
  rely on SDSS redshifts. \xrt{}~151219 also has redshift estimates of
  0.67 \citep{kuijken19} and 0.68 \citep{wright19} using combined data
  from KiDS and VIKING, 0.27 \citep{de_jong17} using only KiDS, and
  0.61 from the Hyper Suprime-Cam Subaru Strategic Program
  \citep{nishizawa20}. Large errors for a small fraction of
  photometric redshift estimates are common, which we believe explains
  the deviating value of 0.27.
\item Our SED fit results in a preliminary redshift of 0.088 for
  \xrt{}~110621. Additionally, it is located 11~arcsec (corresponding
  to a projected distance of 20~kpc at the final redshift) away from
  an $m_r=15.6$~\magab{} galaxy and 44~arcsec (77~kpc) from an
  $m_r=13.4$~\magab{} galaxy (Figure~\ref{fig:res}). The former has a
  photometric redshift of 0.097 \citep{dalya16} and the latter a
  spectroscopic redshift of 0.095 \citep{jones09}. For these reasons,
  we adopt $z=0.095$ as the final value for \xrt{}~110621.
\item \xrt{}~030206 is well-fit by SB10 in the COSMOS sample, which is
  the second bluest starburst in the sample. Furthermore, the SED
  captures the Balmer break well, which no other SED type does
  adequately. This lends confidence to the photometric redshift and we
  take $z=1.17$ for this galaxy.
\item \xrt{}~070618 has no detected host but is located 11~arcsec
  (58~kpc) away from an $m_r=20.83$~\magab{} galaxy and 21~arcsec
  (107~kpc) from an $m_r=18.8$~\magab{} galaxy
  (Figure~\ref{fig:res}). We assume that all three sources are located
  at the same redshift and perform SED fits to the two bright
  galaxies. Reasonable agreement is achieved for $z=0.37$, which we
  take as the estimate for \xrt{}~070618 despite the uncertainties.
\item \xrt{}~060207 lacks information for any type of redshift
  estimate. Therefore, we simply adopt a fiducial value of $z=0.3$.
  This redshift results in reasonable inferred SBO parameters and is
  also motivated by the photometry limits on a host galaxy. A much
  lower redshift would require a very faint host, whereas a much
  higher redshift would imply an extreme \xray{} transient. However,
  we stress that the redshift of this transient is clearly highly
  uncertain.
\item \xrt{}~100424 has acceptable SED fits for redshifts in the range
  0.08--0.4. There are two galaxies with $m_r=16.0$ and $16.5$,
  respectively, at distances of 63~arcsec (146~kpc) at $z=0.10$ and
  71~arcsec (166~kpc) at $z=0.16$. These two redshifts are photometric
  redshifts from the GLADE catalog \citep{dalya16}. We adopt a
  redshift of 0.13 for \xrt{}~100424 (implicitly for all three
  galaxies) based on the combined information.
\item \xrt{}~151128 has acceptable SED fits for redshifts in the range
  0.2--0.65. The best fit is around 0.48, which we adopt as the
  favored redshift.
\item We also use the fiducial value of $z=0.3$ for \xrt{}~050925 and
  160220. In contrast to \xrt{}~060207, there are NIR detections for
  these sources and $z=0.3$ results in reasonable host NIR absolute
  magnitudes.
\item The redshift for \xrt{}~140811 is solely based on SED fits,
  which favor a redshift of 0.57.
\item The redshift for \xrt{}~040610 is determined based on
  photometric redshifts from SDSS12 \citep{alam15} of eight
  neighboring galaxies within 20~arcsec (123~kpc). The average
  redshift is 0.5 with a standard deviation of 0.17. We take $z=0.5$
  as the redshift of \xrt{}~040610 despite the large uncertainties.
\end{itemize}

\section{Observations and Data Reduction}\label{sec:obs}
We perform data reduction of all selected sources
(Table~\ref{tab:identifiers}) largely following standard
procedures. This data reduction is independent of the automatic
processing used for finding the transients
(Appendix~\ref{app:search}). We use \xmm{} Science Analysis System
(SAS) version 18.0.0 \citep{gabriel04} and the latest CCF as of 2019
December 11. We inspect the images from the Optical Monitor
\citep{mason01} on board \xmm{} when available, but find no variable
emission that appears connected with the \xray{}
transients. Henceforth, we only consider data from the European Photon
Imaging Camera (EPIC). All 12 sources are covered by pn and MOS2,
while 8 sources are covered by MOS1 since part of its FoV has been
disabled.

We apply the latest calibration using the tasks \texttt{epproc} and
\texttt{emproc}. We also apply filters to keep data only in the range
0.3--10~keV with standard filtering parameters. We select temporary
source regions and final background regions by hand. The background
regions typically have radii of ${\sim}$1~arcmin and are selected to
be close to the source, free of other sources, and on the same CCD
chip as the source. We use these source and background regions to
construct preliminary light curves.

Next, we define five time intervals: before, during, after, and the
first and second halves of the transient. The halves of the transient
are defined to have approximately equal fluences. These two halves are
only used for time-resolved spectral analysis
(Section~\ref{sec:fit}). All time intervals are defined by eye based
on inspections of the preliminary source light curves.

We remove periods of high background before and after the defined
duration of the transient following standard procedures. The
background levels during the transients are not high enough to
significantly affect the analysis. Only \xrt{}~110621 occurs during a
high background interval, but is still clearly detected with a
signal-to-noise ratio (S/N) of 5. The median S/N across all sources is
7, where the longer transients contribute to a lower ratio. We note
that the S/N only includes the photon number statistics and does not
represent the detection statistic, which also includes the spatial
information. The lowest detection statistic as defined by
\texttt{emldetect} \citep{watson09} is 20 for \xrt{}~151128. This
approximately corresponds to a false detection probability of
$2\times{}10^{-9}$.

With the time intervals, it is possible to obtain an improved estimate
of the source position by fitting to the image created from the
duration of the transient. We use the SAS task \texttt{edetect\_chain}
to fit for the position in the 0.3--10~keV range in all cameras
simultaneously. We add in quadrature a systematic uncertainty of
1.2~arcsec\footnote{\xmm{} Calibration Technical Note
  XMM-SOC-CAL-TN-0018} to the statistical uncertainty from
\texttt{edetect\_chain}, to obtain the final position uncertainty. We
adopt this position as the final source position and create the final
source regions. The radii for these regions are chosen to maximize the
S/N, which is computed by the SAS task \texttt{eregionanalyse}.

The next step of the data preparation is to create a light curve using
the updated source region, and create images and extract spectra
during each of the five time intervals.  The light curve and images
are created following standard \xmm{} data reduction procedures. We
largely follow standards when extracting the spectra as well. However,
we group all source spectra to at least 1 count per spectral bin,
which is necessary for using $C$-stat in
XSPEC\footnote{\url{https://heasarc.gsfc.nasa.gov/xanadu/xspec/manual/XSappendixStatistics.html}}.
The ancillary response files (ARFs) and response matrix files (RMFs)
for the individual cameras are generated using the SAS tasks
\texttt{arfgen} and \texttt{rmfgen}, respectively. Solely for
presentation purposes, we also combine the spectra from the three
cameras using \texttt{especcombine}. The individual spectra used for
the merged EPIC spectra are binned using a common channel grid
corresponding to uniform bins of width
0.2~keV.

To determine upper limits on the source flux during the time intervals
before and after the transient, we use the SAS task
\texttt{esensmap}. The task computes the photon flux required for a
hypothetical source to be detected in the combination of all three
instruments. To convert the photon flux to physical units, we use the
best-fit model of the time-averaged spectrum
(Section~\ref{sec:fit}). For the upper limits, we choose a detection
threshold (\texttt{DET\_ML}) of 5.9, which corresponds to a chance
probability for detection of $2.7\times{}10^{-3}$ ($3\sigma{}$) under
the null hypothesis.

To estimate the peak flux of each transient, we start by finding the
shortest time interval during which 25\,\% of all source photons are
detected. This is done by searching a combined EPIC source event
list. Having determined the time interval with the highest photon
flux, we use \texttt{epiclccorr} to apply all corrections to obtain
the corrected count rate with error bars. This task is usually used to
produce light curves, whereas we use it here to compute a count rate
during a single time interval. This is performed on the cameras
individually, with the single-bin time interval provided as input. The
final step is to convert the peak count rates from \texttt{epiclccorr}
to fluxes. We do this by scaling the fluxes obtained from the
time-integrated spectral fits (Section~\ref{sec:fit}) by the ratio of
the peak to the time-averaged photon count rate.

When the flux is high, it is possible that two incident, spatially
close photons are interpreted incorrectly. They could be detected as a
single photon with higher energy or flagged as bad, a phenomenon
referred to as pile-up. There is some potential for pile-up during the
brightest phases of our transients. \xrt{}~070618 is the most likely
to suffer from pile-up, with a peak count rate of 0.4 photons per
frame in the pn CCD. The standard SAS method used for assessing
pile-up, \texttt{epatplot}, is inconclusive because too few photons
are detected during the short transient. Instead, we manually inspect
the raw (ODF), calibrated (unfiltered), and final event lists and find
no signs of suspected pile-up events. All light curves are practically
identical, including the marginal double-peaked shape
(Section~\ref{sec:observed}).

Finally, we comment on the flux of the object \xrt{}~151219. It is
close to (but not on) a chip gap on the pn CCD. A part of the standard
procedure when creating pn spectra is to reject events close to chip
gaps (\texttt{FLAG==0}). This is performed because of energy
calibration uncertainties for events not completely contained on the
chip. For this object, this means rejecting ${\sim}$20\,\% of the pn
events. The spectral shape is unaffected, but the overall flux is
likely underestimated by ${\sim}$15\,\% (the source is covered by both
pn and MOS2). This does not affect any conclusions qualitatively and
we do not attempt to correct for it.

\section{Spectral Fitting}\label{sec:fit}
We fit two simple models to the data: a blackbody (\texttt{zbbody})
and power-law (\texttt{zpowerlw}) model. Both models also include an
absorption component for the Milky Way absorption (\texttt{tbabs},
\citealt{wilms00}) and a redshifted absorption component
(\texttt{ztbabs}) representing host galaxy absorption. The Milky Way
column density is frozen to the weighted estimate of
\citet{willingale13}, whereas the redshifted column densities are
fitted for. Both absorption components use the abundances of
\citet{wilms00} and cross sections of \citet{verner95}.

We freeze the redshifts to our estimated values
(Table~\ref{tab:host_par}) and ignore the uncertainties that are
implicitly introduced. The redshifts only have a minor impact on the
fitted parameters (Section~\ref{sec:observed}) but have significant
implications for the inferred physical properties, which are discussed
separately (Section~\ref{sec:sn_prop}).

The blackbody model is a simplified representation of a complete
physical model for the time-integrated spectrum of a SN SBO
(Section~\ref{sec:sbo_mod}). The integrated spectra of SBOs are not
expected to be perfectly thermal, but a blackbody is a reasonable
approximation given the large uncertainties. \citet{sapir13} showed
that the time-integrated spectrum of a SBO peaks at
$h\nu_\mathrm{peak} = 3T_\mathrm{peak}$, where $h\nu_\mathrm{peak}$ is
the photon energy at the spectral peak (measured in fluence per
logarithmic frequency; $\nu{}F_\nu$) and $T_\mathrm{peak}$ is the peak
surface temperature. In comparison, the spectral peak of a blackbody
of temperature $T$ peaks at $h\nu_\mathrm{peak}=2.82T$.  Therefore,
the fitted temperature is a reasonable approximation of the peak
surface temperature.

The interpretation of the power-law model is simpler than the
interpretation of the blackbody because we only use the power law as a
phenomenological model. Its purpose is to quantify the observed
spectral slope by fitting for the photon index ($\Gamma$). Moreover,
the goodness of fit is also used for comparisons with the blackbody
model.

We also fit the blackbody and power-law models to the spectra from the
first and second halves of the transients. However, since the aim is
to solely quantify spectral evolution, the host absorptions are frozen
to the best-fit values from the time-integrated fits.  This assumes
that all information about the evolution of the spectral shape is
captured by the temperature or photon index.

\subsection{Technical Aspects}\label{sec:tech}
We use XSPEC version 12.10.1f \citep{arnaud96} for the spectral
analysis. The spectra are fitted simultaneously to data from all three
EPIC CCDs in the entire 0.3--10~keV range. We do not introduce a free
cross-normalization between the instruments. The relative calibration
of the instruments are accurate to within 5--10\,\% \citep{madsen17,
  plucinsky17}, but additional free constants between the instruments
cannot be reliably constrained in a few cases of extremely few counts.
The analysis is performed with the $C$-stat fit statistic
\citep{cash79} in XSPEC, which is the $W$~statistic when a Poissonian
background is
included\footnote{\url{https://heasarc.gsfc.nasa.gov/xanadu/xspec/manual/XSappendixStatistics.html}}.

Confidence intervals for the parameter estimates are computed using
the \texttt{error} command in XSPEC, which varies the fit parameters
until the change in fit statistic reaches a given threshold. This
assumes that the difference in fit statistic is distributed as
$\chi^2_q$ for $q$ degrees of freedom. This is a reasonable
approximation even for Poissonian data except for a small additive
correction $\propto n^{-1/2}$ for $n$ photons \citep{cash79,
  yaqoob98}. For the time-integrated fits, there are three free
parameters. An appropriate threshold for 1$\sigma$ confidence
intervals and three independent parameters is a change in fit
statistic of 3.5 \citep{avni76, lampton76, cash76}. Analogously, for
the time-resolved fits with two free parameters, the corresponding
value is 2.3. For reference, we note that X-ray analyses commonly use
a critical value of 2.706, which represents the 90\,\% interval for
one parameter of interest.

The goodness of fits are computed by simulating 1000 spectra using the
\texttt{fakeit} command in XSPEC. The spectra are simulated using the
best-fit model at the energy resolution of the channels of the
instruments. The faked spectra are then grouped to at least 1 count
per bin, analogously to the real data. The models are then fitted to
these spectra. The final goodness measure is represented by the
fraction of fits to faked spectra with better test statistic than fits
to the observed data. A good fit should on average result in a
fraction of 0.5. For our purposes and given the systematic
uncertainties, we consider goodness fractions relatively close to 1 as
acceptable. The fit statistics and goodness measures for all fits are
provided in Table~\ref{tab:stat}.

\section{Observed Properties}\label{sec:observed}
\begin{figure*}
  \centering
  \includegraphics[width=0.32\textwidth]{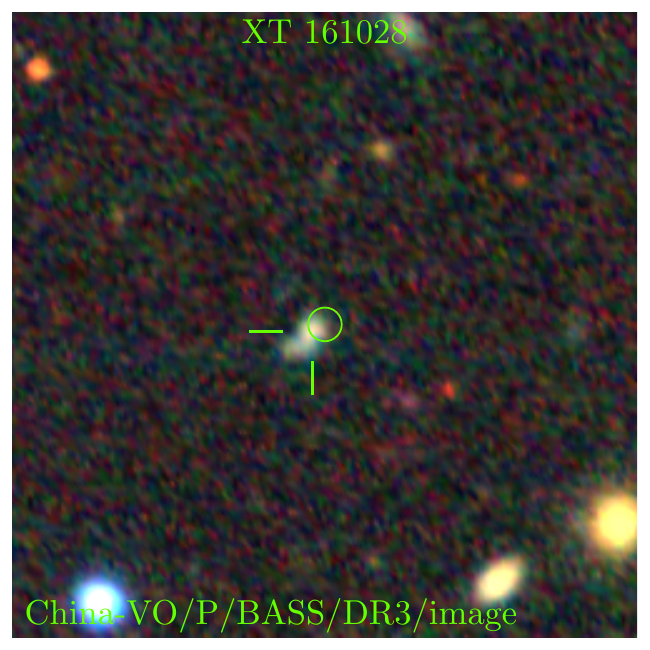}
  \includegraphics[width=0.32\textwidth]{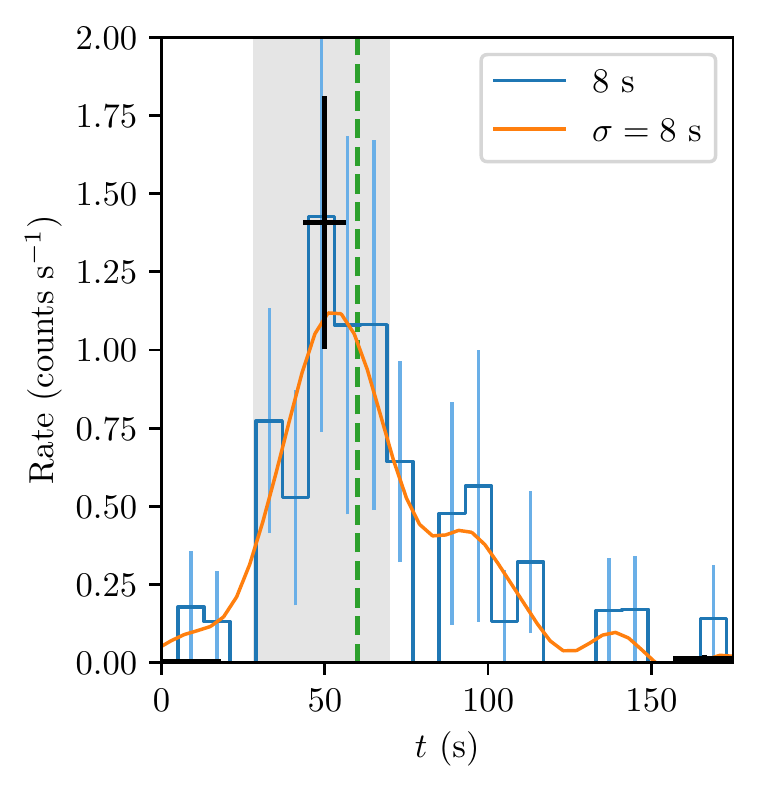}
  \includegraphics[width=0.32\textwidth]{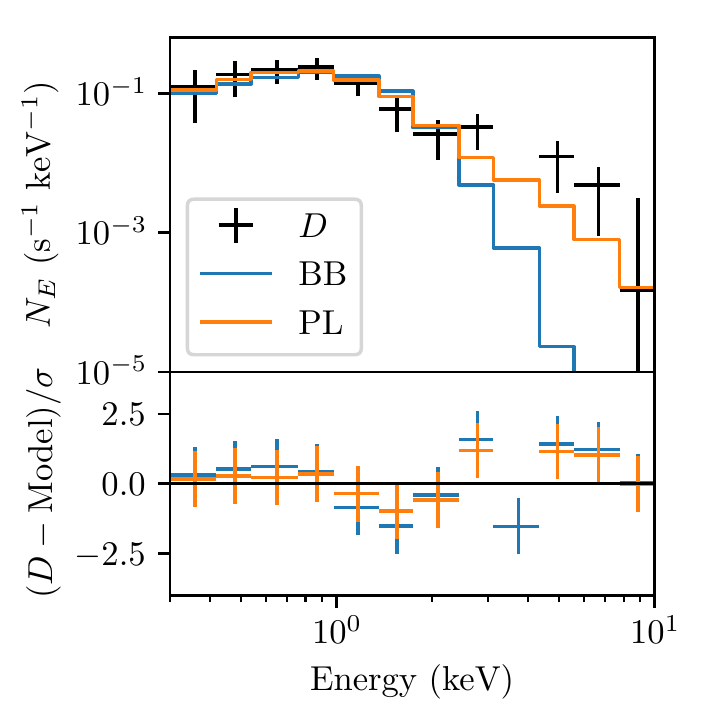}

  \includegraphics[width=0.32\textwidth]{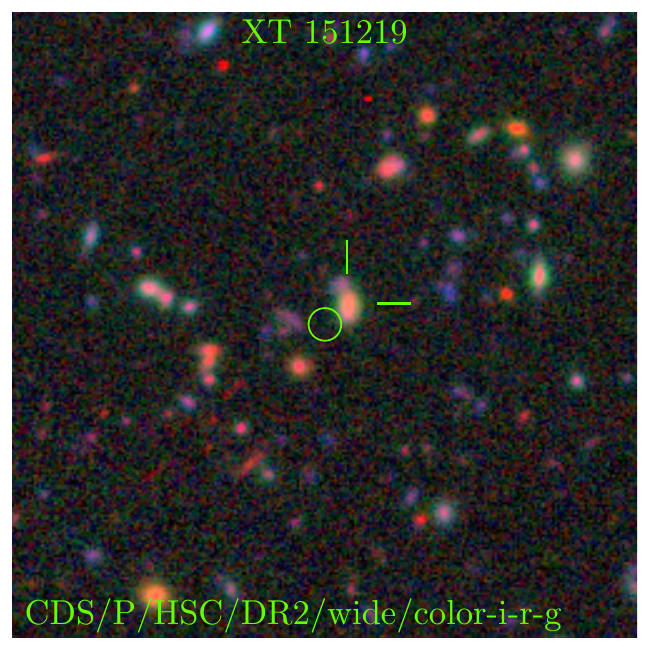}
  \includegraphics[width=0.32\textwidth]{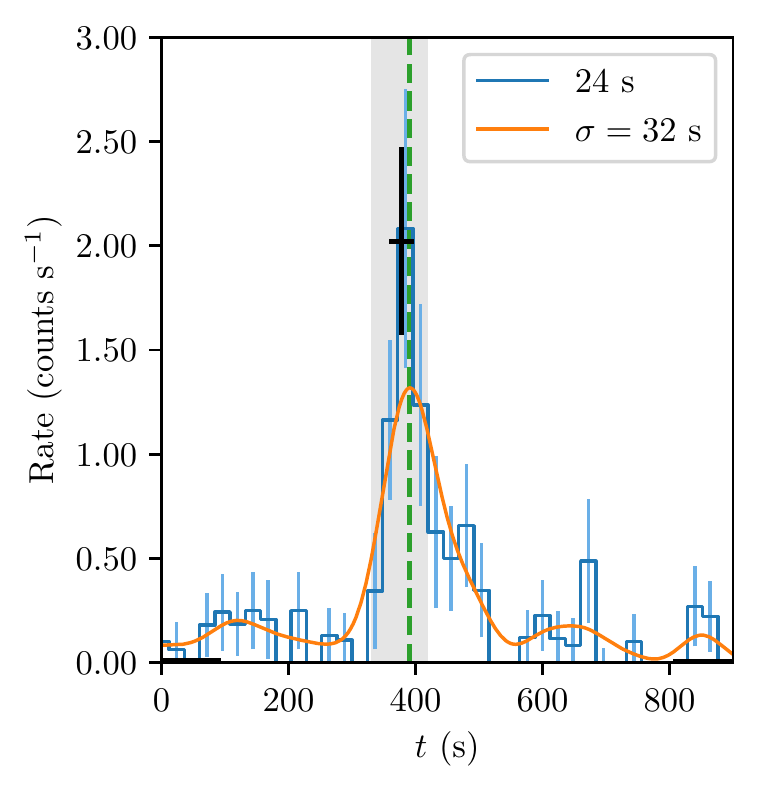}
  \includegraphics[width=0.32\textwidth]{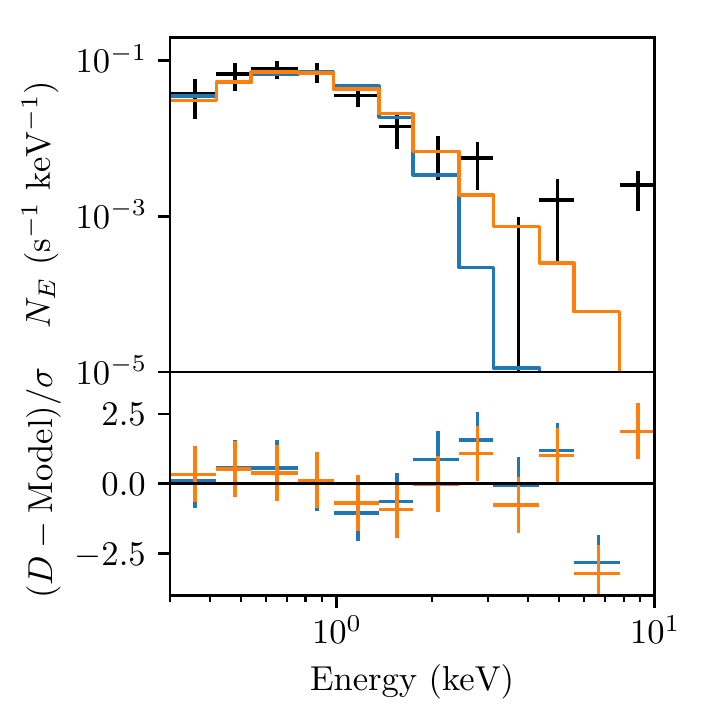}

  \includegraphics[width=0.32\textwidth]{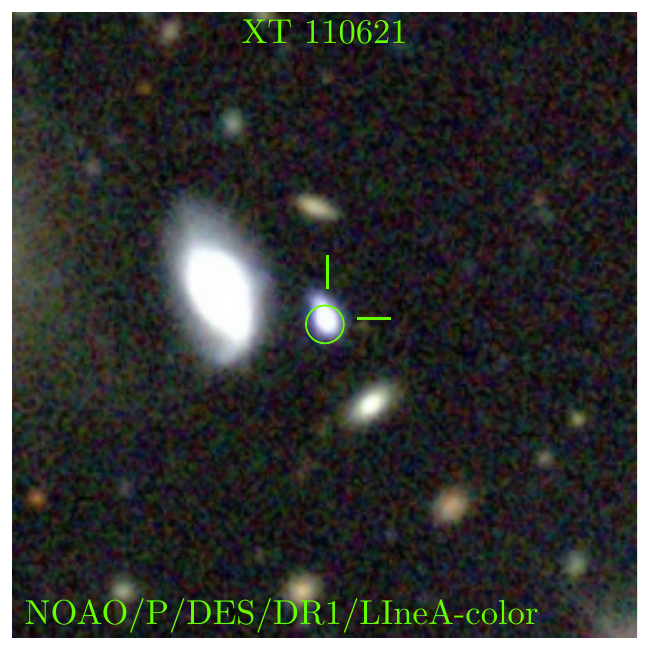}
  \includegraphics[width=0.32\textwidth]{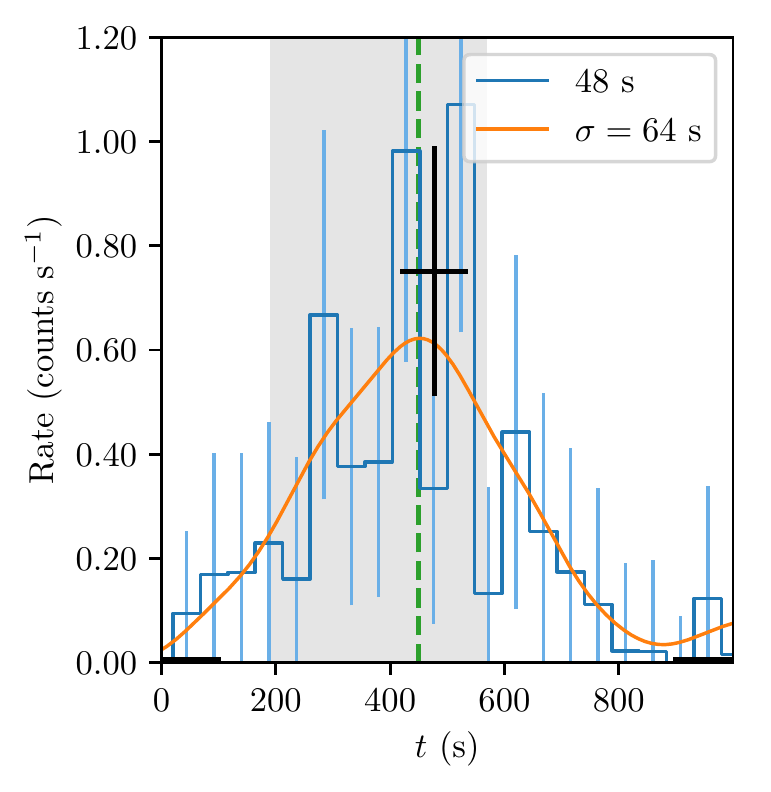}
  \includegraphics[width=0.32\textwidth]{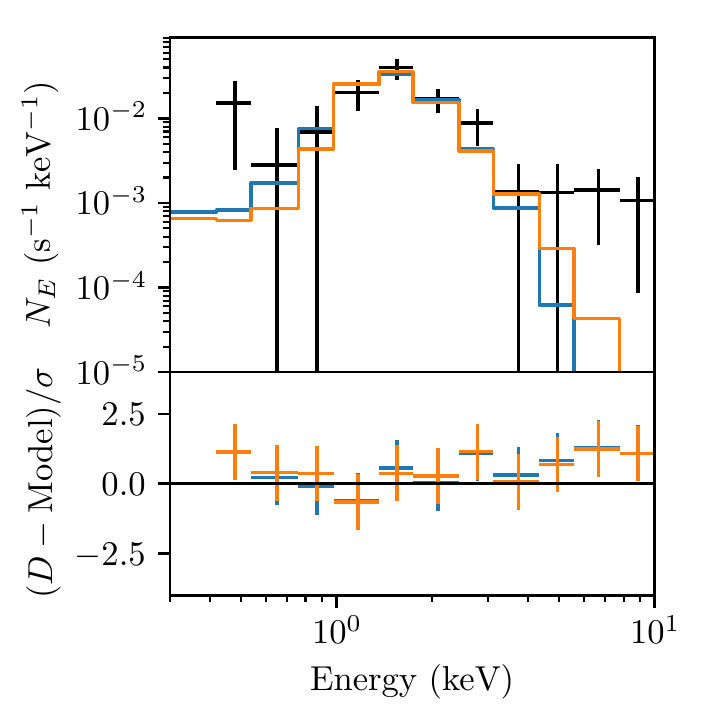}

  \caption{\textit{Continues on subsequent pages (1/4).} SBO
      candidates in descending order of confidence of being a SN SBO.
    Left column: Sky images centered on the positions of the \xray{}
    transients (green circles with 1$\sigma$ radii). The green lines
    mark the candidate host galaxies. The bottom texts are the sources
    of the images, which are all from HiPS except for VISTA and Subaru
    data. The color scales are arbitrary and are chosen to maximize
    contrast. The image colors loosely represent source colors for
    multi-band data. All images are $1\times{}1$~arcmin$^2$ and the
    lines pointing to the galaxies are 3~arcsec. North is up and east
    is left. Middle column: Combined EPIC light curves in the observed
    0.3--10~keV range. Blue lines are binned light curves with light
    blue uncertainties, and orange lines are smoothed light curves
    (bin sizes and kernel widths are given in the legends). The dashed
    green lines show the separation into first and second halves for
    the time-resolved spectral analysis. The gray regions show the
    typical timescales ($t_{R/c}$; Section~\ref{sec:sbo_mod}). The
    black crosses are the peak fluxes (Section~\ref{sec:obs}) and the
    horizontal black lines (lower left and right corners, barely
    visible in some cases) are the flux limits
    (Table~\ref{tab:lc}). Right column: Upper segments show the
    observed time-integrated spectra (black crosses) fitted with the
    absorbed blackbody (blue) and power-law (orange)
    models. Lower segments show the residuals normalized by the
    uncertainties using the same colors as above for the respective
    models. For visual clarity, we combine all EPIC data and bin to
    12 energy bins logarithmically.
    \label{fig:res}}
  
\end{figure*}

\addtocounter{figure}{-1}
\begin{figure*}
  \includegraphics[width=0.32\textwidth]{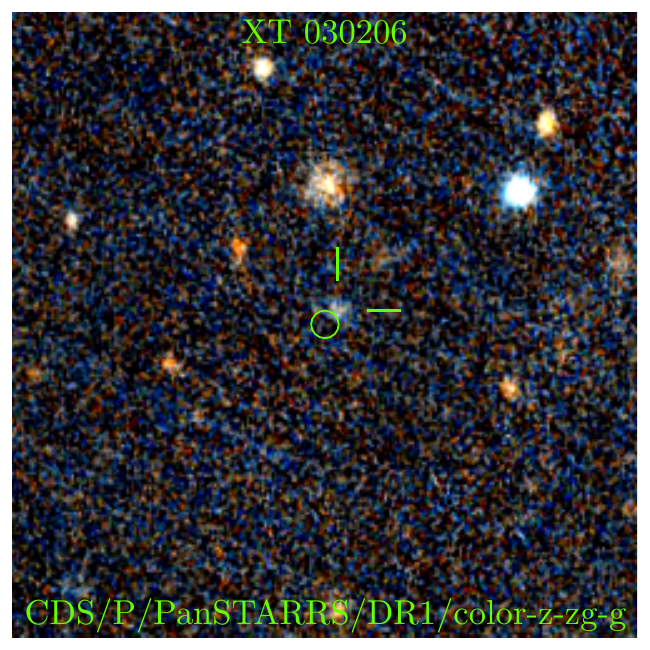}
  \includegraphics[width=0.32\textwidth]{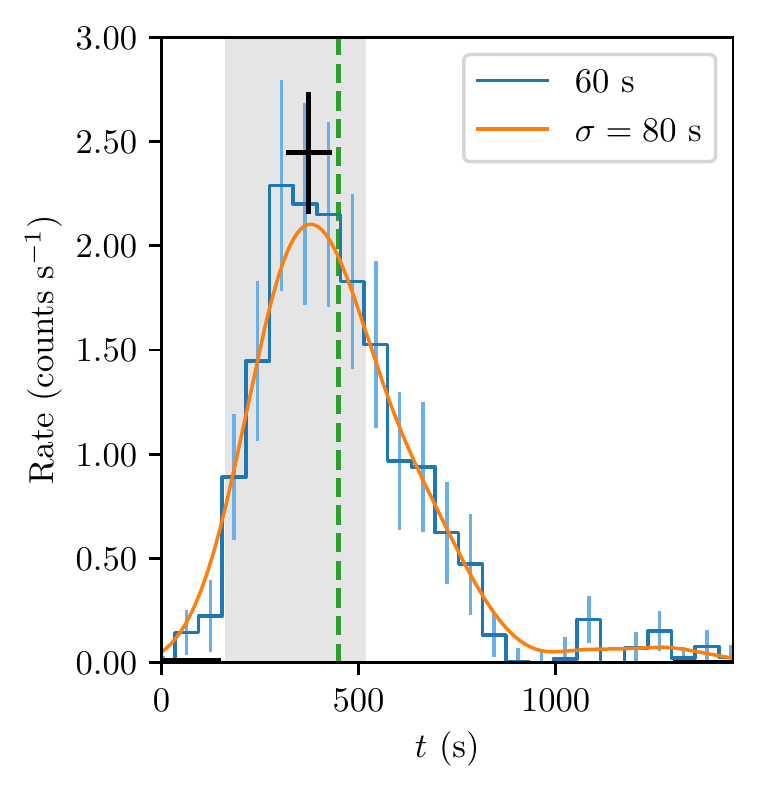}
  \includegraphics[width=0.32\textwidth]{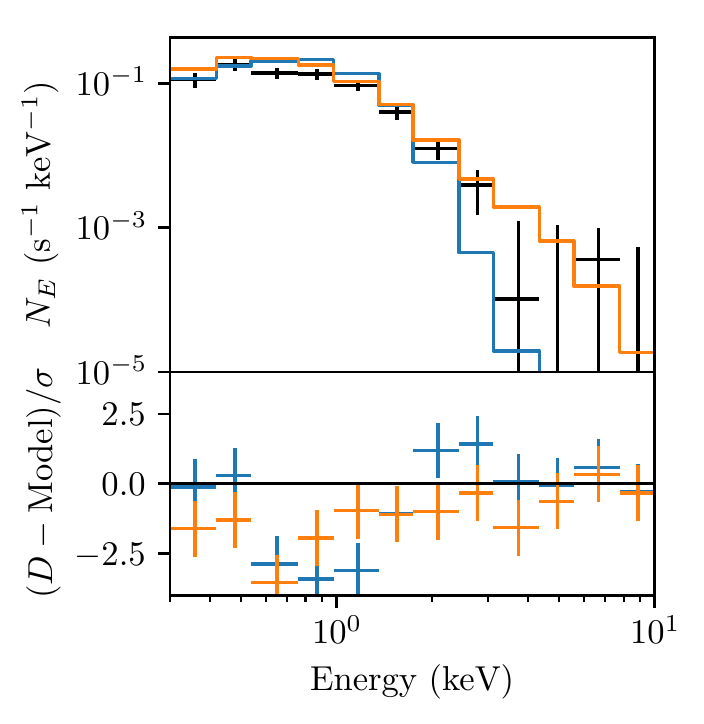}

  \includegraphics[width=0.32\textwidth]{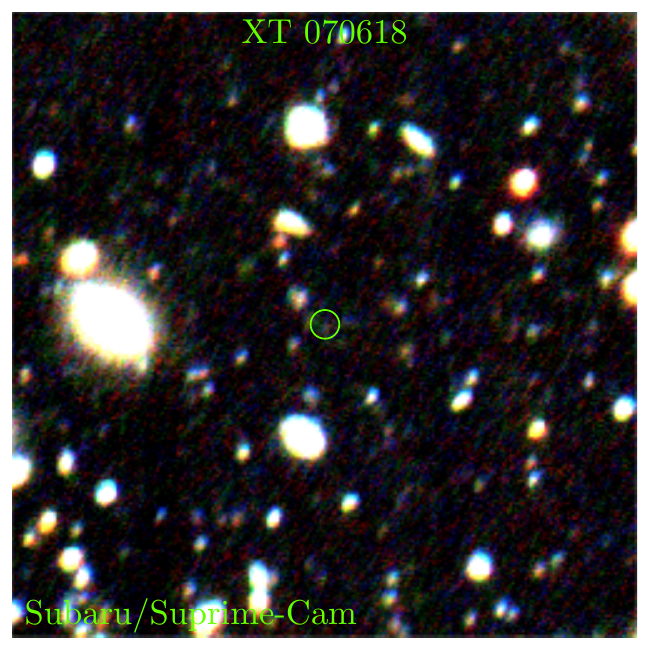}
  \includegraphics[width=0.32\textwidth]{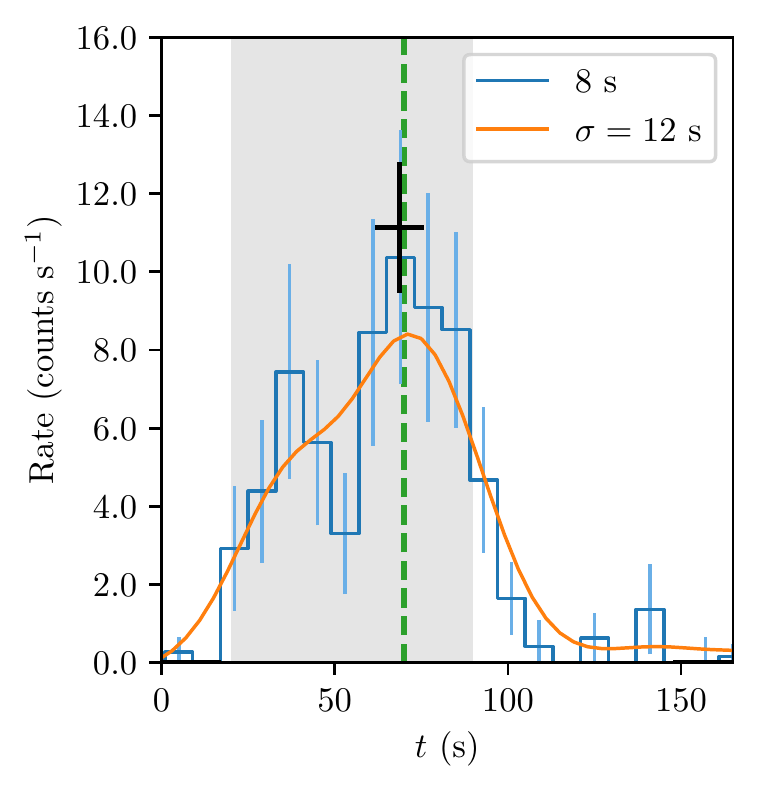}
  \includegraphics[width=0.32\textwidth]{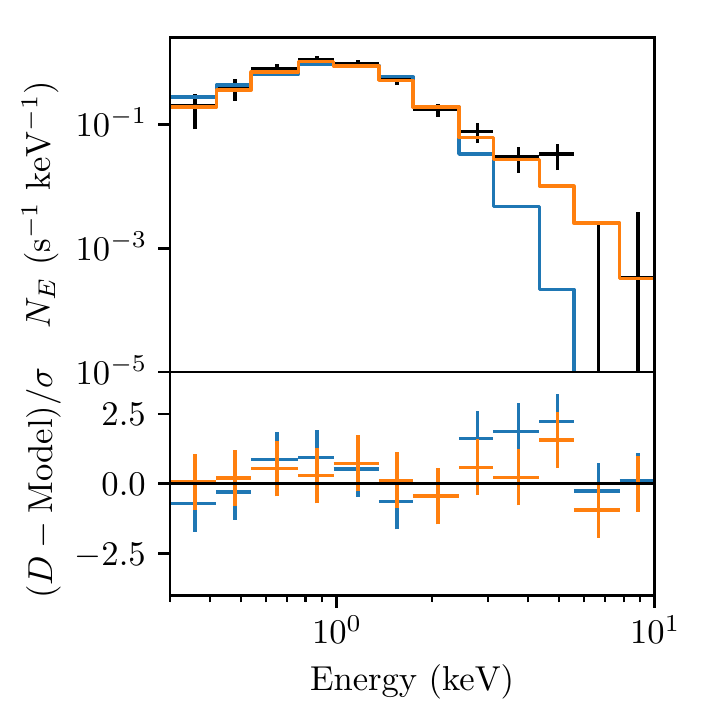}

  \includegraphics[width=0.32\textwidth]{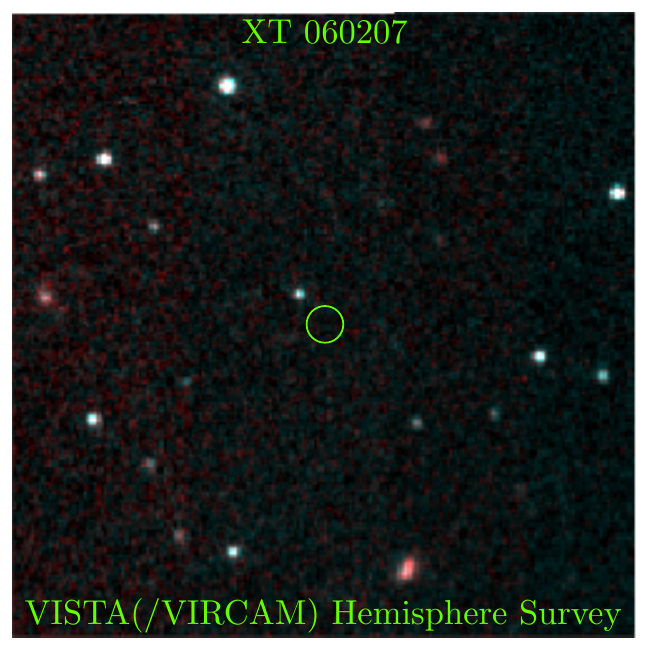}
  \includegraphics[width=0.32\textwidth]{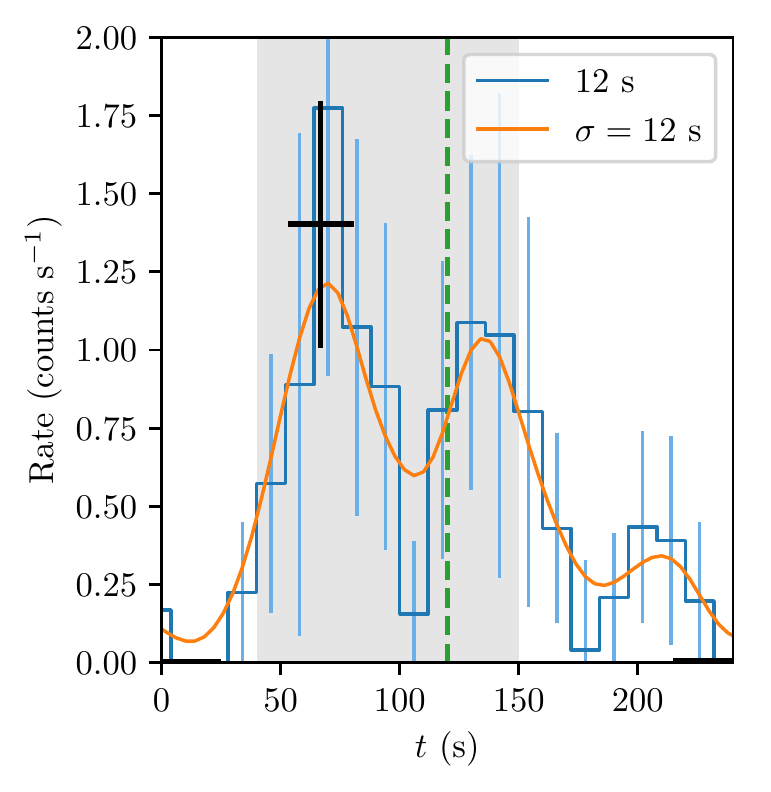}
  \includegraphics[width=0.32\textwidth]{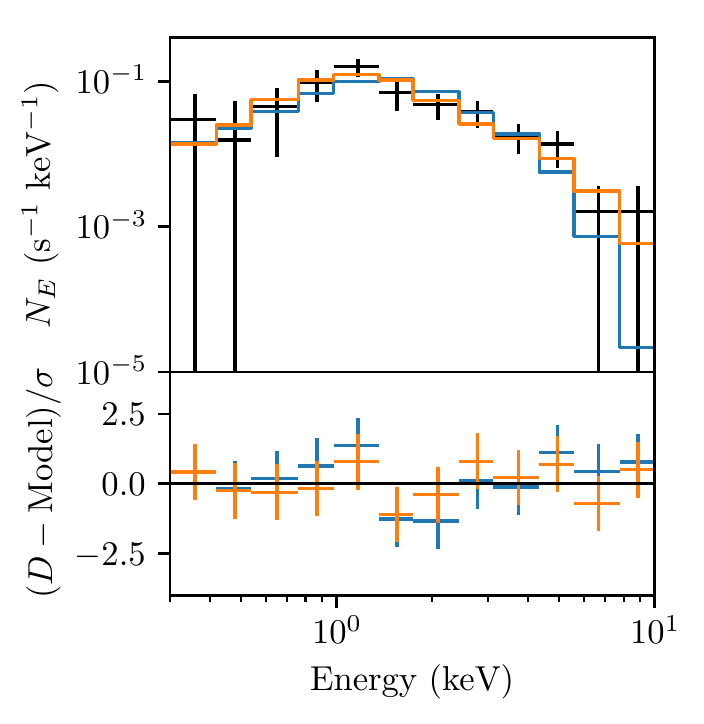}
  \caption{\it Continued (2/4).}
\end{figure*}

\addtocounter{figure}{-1}
\begin{figure*}
  \includegraphics[width=0.32\textwidth]{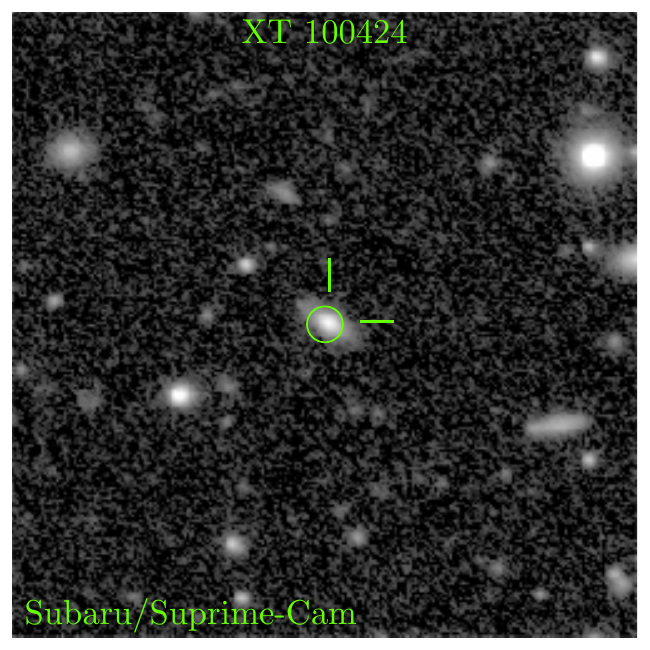}
  \includegraphics[width=0.32\textwidth]{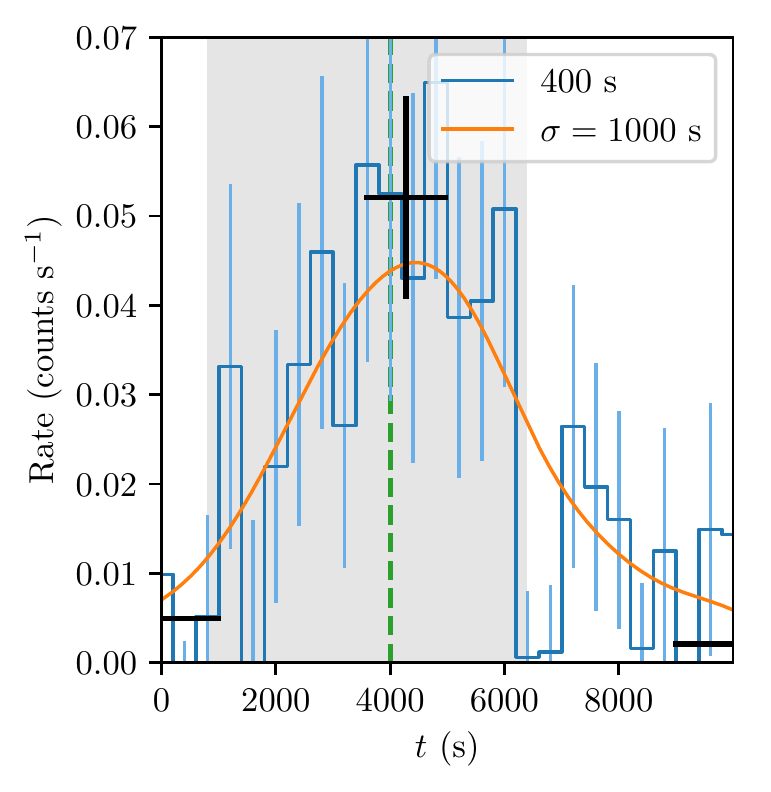}
  \includegraphics[width=0.32\textwidth]{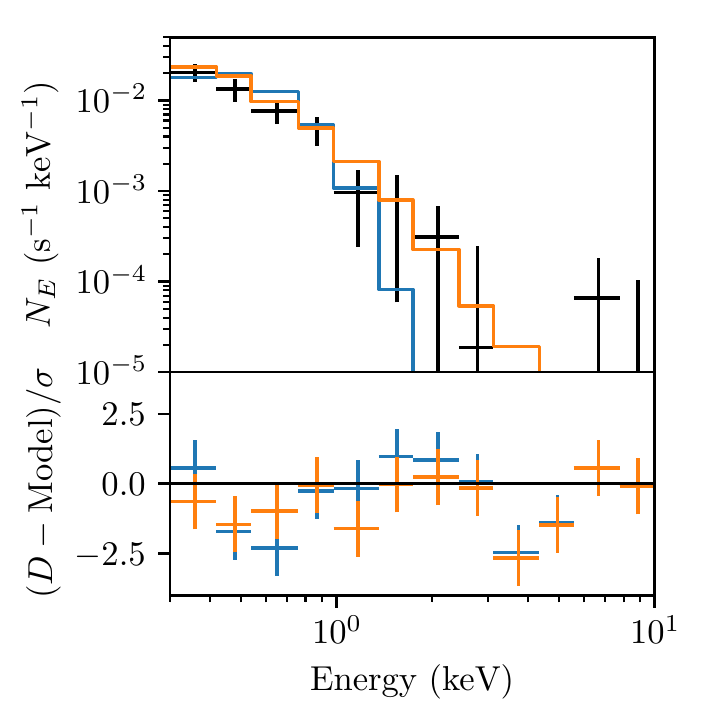}

  \includegraphics[width=0.32\textwidth]{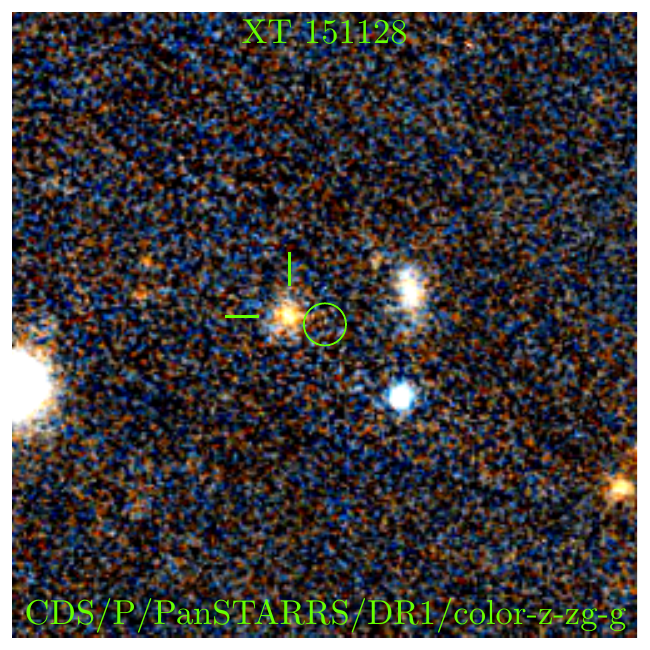}
  \includegraphics[width=0.32\textwidth]{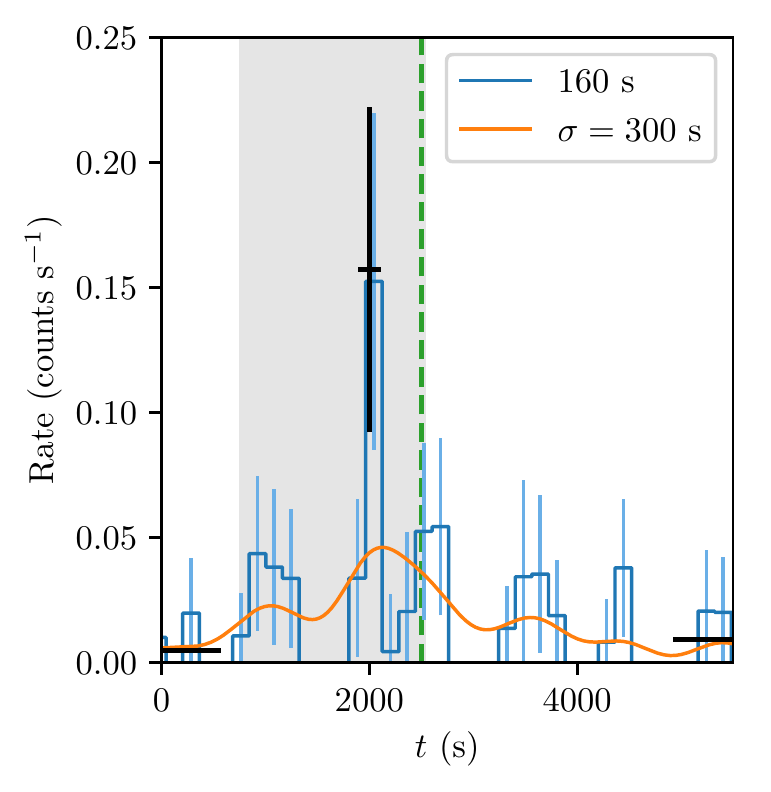}
  \includegraphics[width=0.32\textwidth]{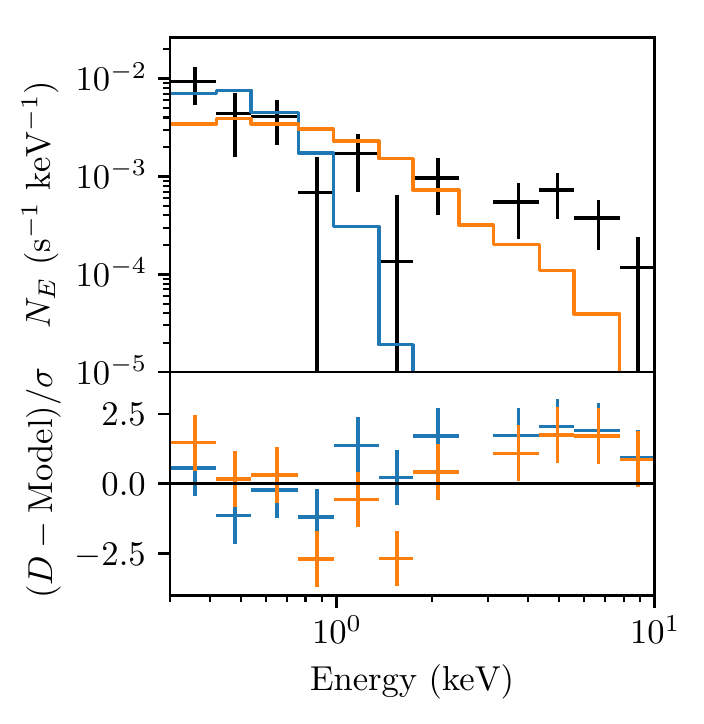}

  \includegraphics[width=0.32\textwidth]{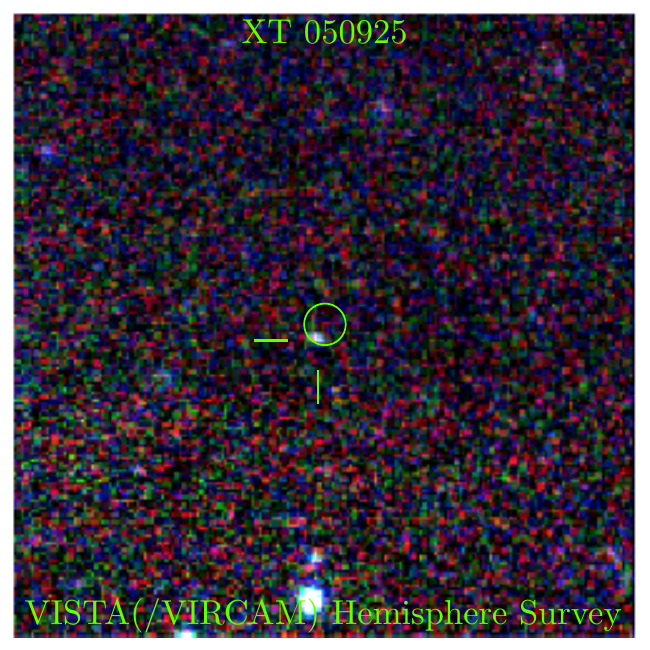}
  \includegraphics[width=0.32\textwidth]{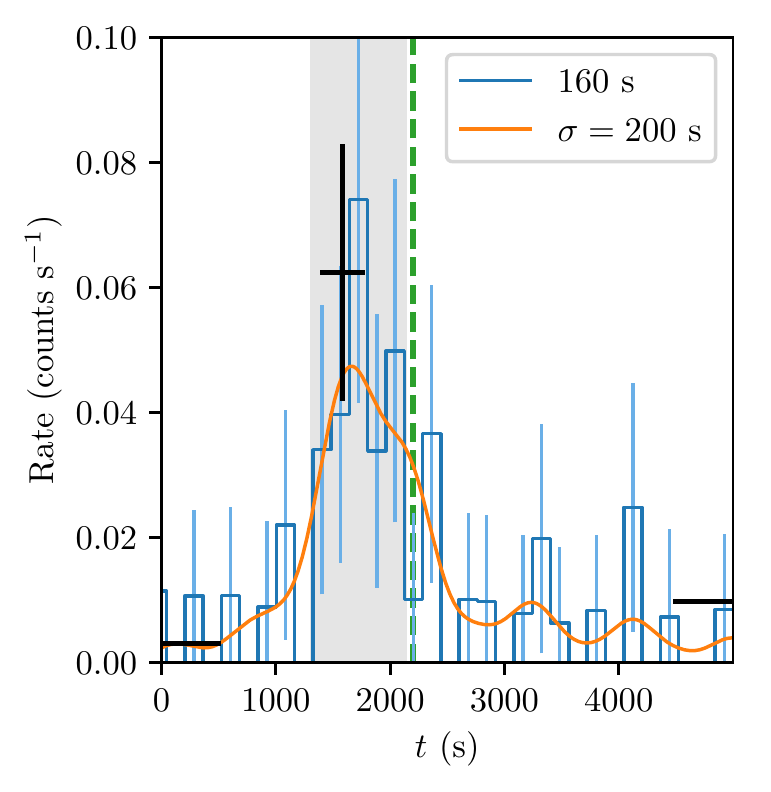}
  \includegraphics[width=0.32\textwidth]{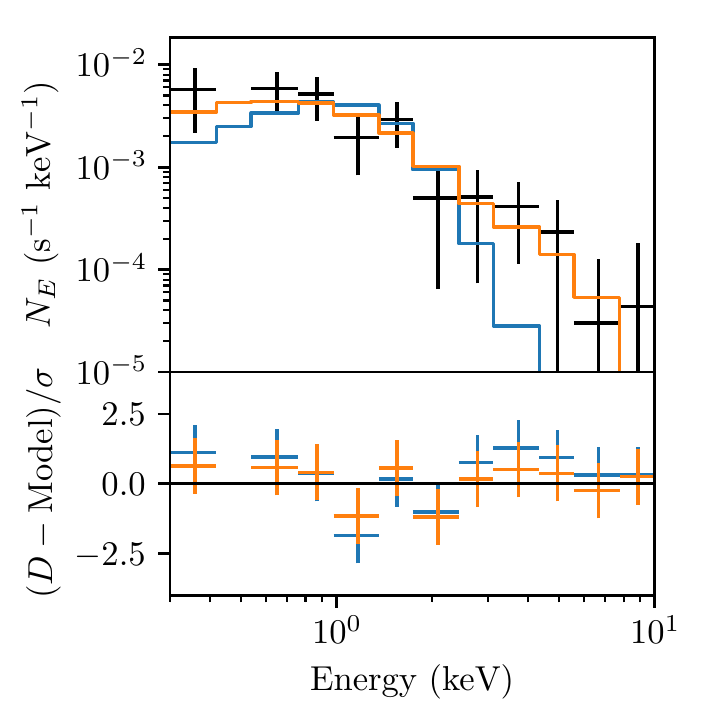}
  \caption{\it Continued (3/4).}
\end{figure*}

\addtocounter{figure}{-1}
\begin{figure*}
  \includegraphics[width=0.32\textwidth]{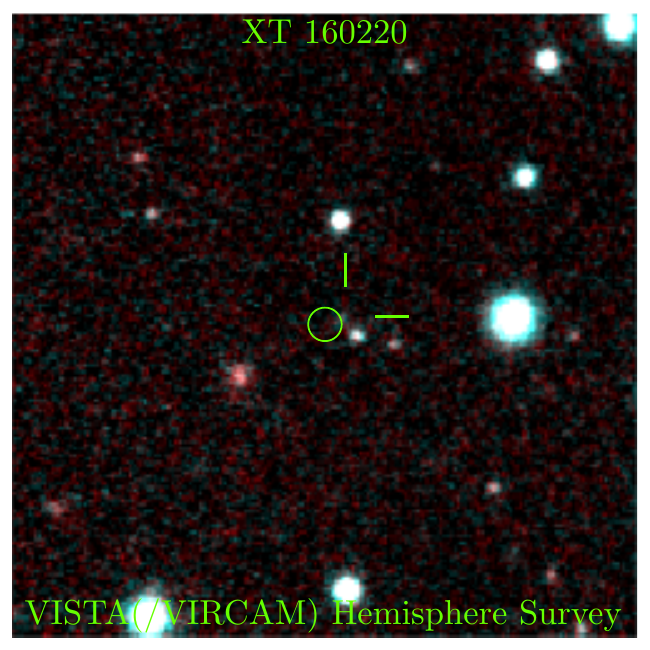}
  \includegraphics[width=0.32\textwidth]{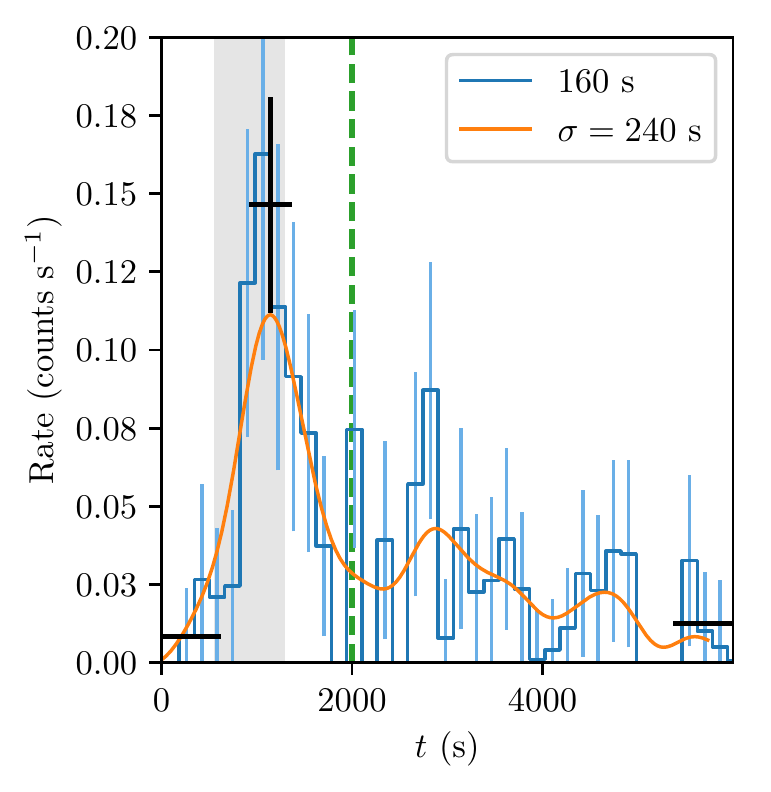}
  \includegraphics[width=0.32\textwidth]{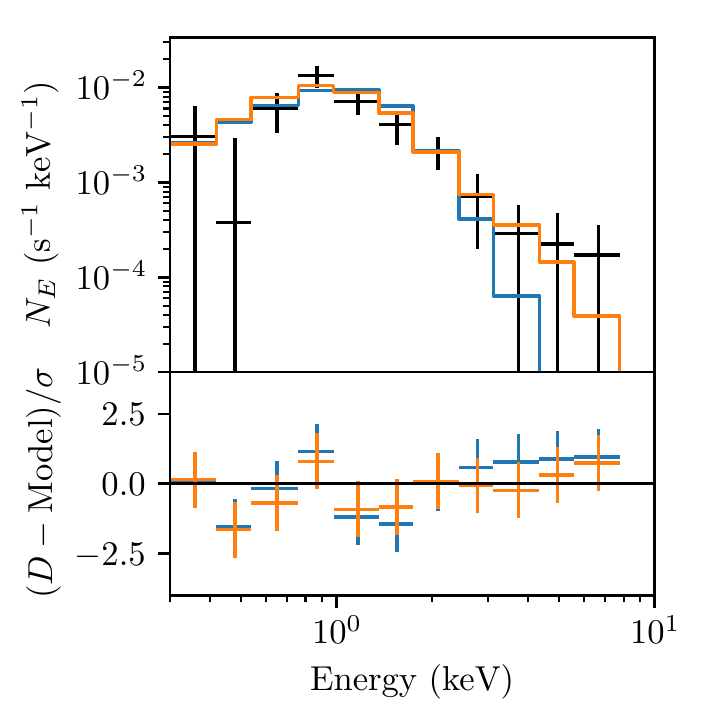}

  \includegraphics[width=0.32\textwidth]{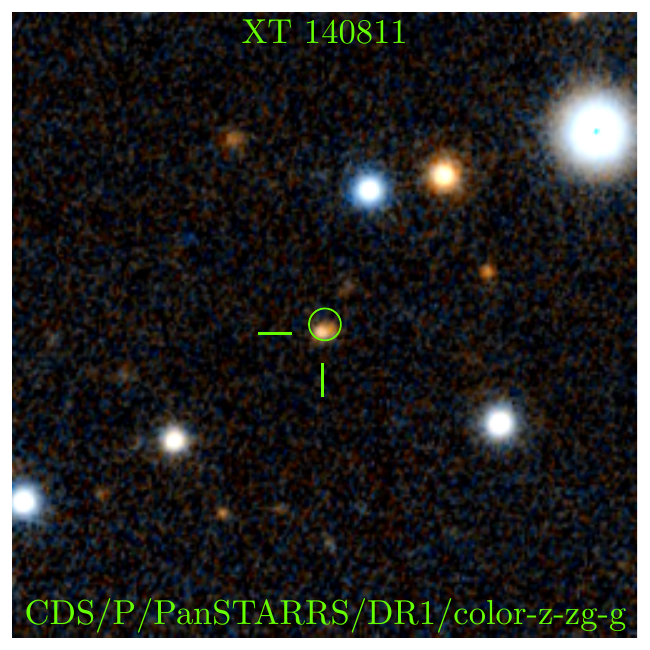}
  \includegraphics[width=0.32\textwidth]{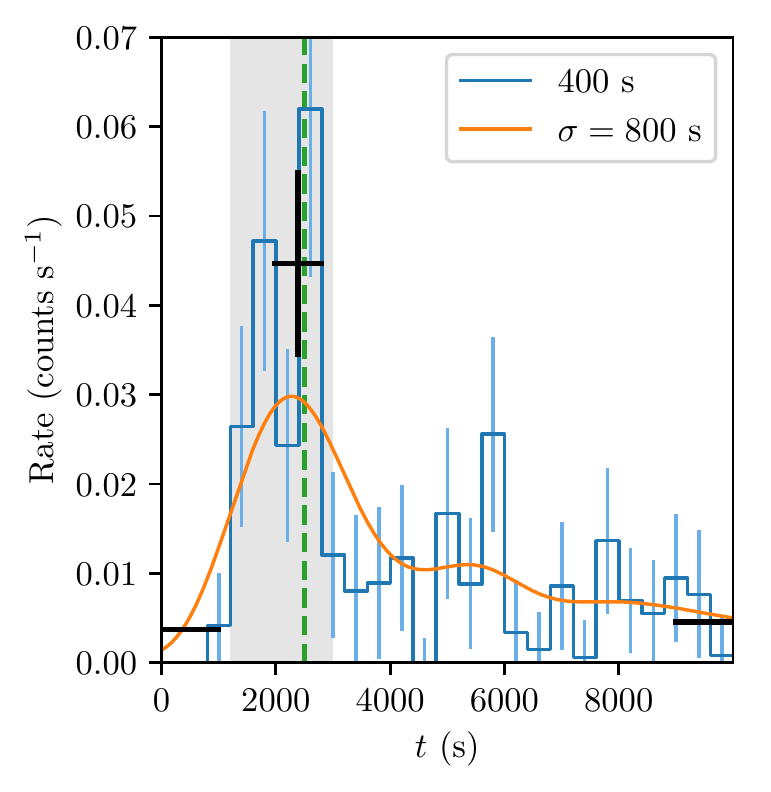}
  \includegraphics[width=0.32\textwidth]{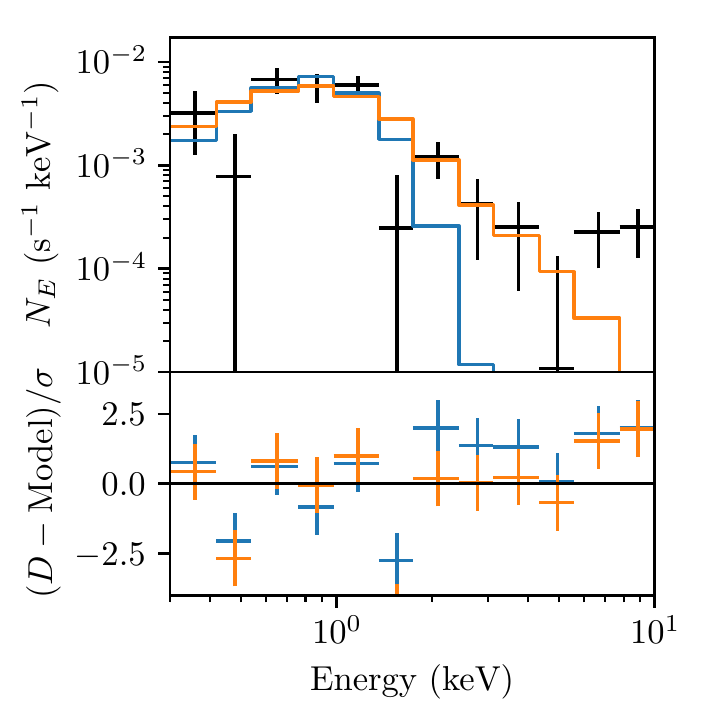}

  \includegraphics[width=0.32\textwidth]{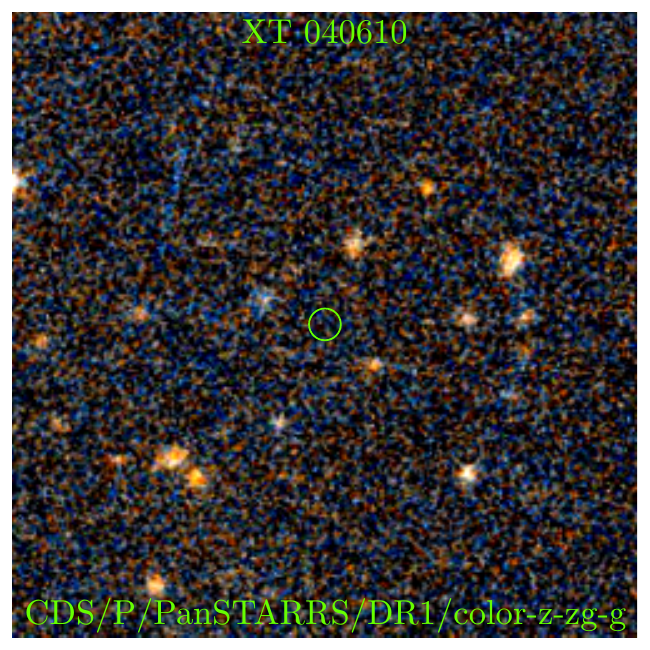}
  \includegraphics[width=0.32\textwidth]{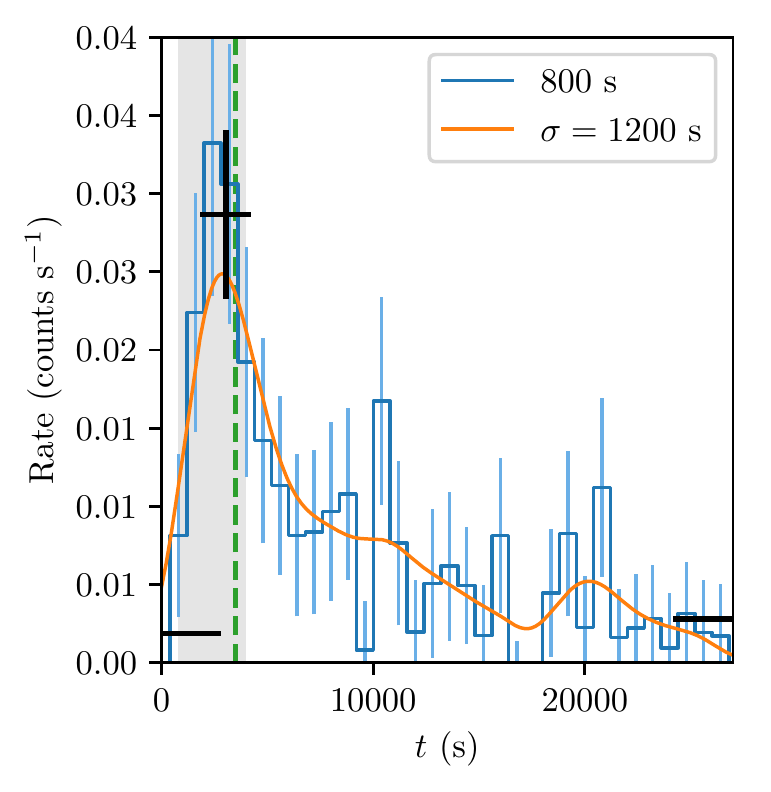}
  \includegraphics[width=0.32\textwidth]{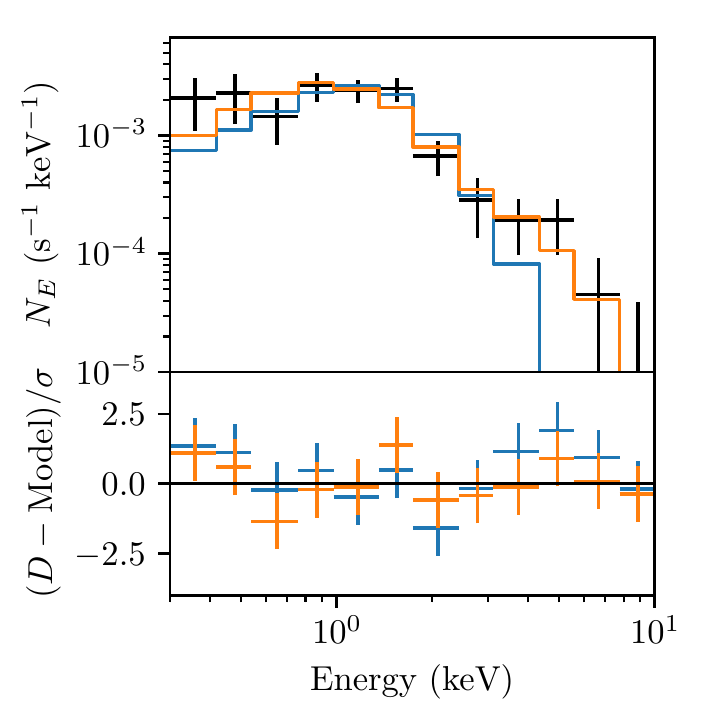}
  \caption{\it Continued (4/4).}
\end{figure*}

Here, we report the observed properties of the SBO candidates. These
are practically independent of any modeling uncertainties. The
redshifts enter into the fitted blackbody temperature and host galaxy
absorption, but do not affect these values much. In
Figure~\ref{fig:res}, we provide the sky images, light curves, and
spectra for all SBO candidates. The purpose of the sky images is only
to show the host galaxy candidates and the neighboring sources
relative to the \xray{} transients. Properties of the light curves and
spectra are presented separately in Sections~\ref{sec:res_lc}
and~\ref{sec:res_spec}, respectively.

\subsection{Light Curves}\label{sec:res_lc}
\begin{deluxetable*}{lccccccccc}
  \caption{Fluxes\label{tab:lc}}
  \tablewidth{0pt}
  \tablehead{
    \colhead{XT} & \colhead{$F_\mathrm{mean}$}                   & \colhead{$F_\mathrm{peak}$}                   & \colhead{$F_\mathrm{before}$}                 & \colhead{$F_\mathrm{after}$}                  & \colhead{$\frac{F_\mathrm{peak}}{F_\mathrm{before}}$} & \colhead{$\frac{F_\mathrm{peak}}{F_\mathrm{after}}$}\\
    \colhead{}   & \colhead{($10^{-13}$~erg~s$^{-1}$~cm$^{-2}$)} & \colhead{($10^{-13}$~erg~s$^{-1}$~cm$^{-2}$)} & \colhead{($10^{-13}$~erg~s$^{-1}$~cm$^{-2}$)} & \colhead{($10^{-13}$~erg~s$^{-1}$~cm$^{-2}$)} & \colhead{}     & \colhead{}} \startdata
  161028         & $3.82\pm0.73$                                 & $15.30\pm4.31$                                & $<0.05$                                       & $0.13\pm0.04$                                 & $>300$         & $100$          \\
  151219         & $2.34\pm0.31$                                 & $17.61\pm3.82$                                & $0.08\pm0.02$                                 & $<0.06$                                       & $200$          & $>300$         \\
  110621         & $2.63\pm0.54$                                 & $7.12\pm2.22$                                 & $<0.06$                                       & $<0.06$                                       & $>100$         & $>100$         \\
  030206         & $5.19\pm0.41$                                 & $18.79\pm2.15$                                & $<0.07$                                       & $<0.02$                                       & $>300$         & $>800$         \\
  070618         & $32.12\pm3.64$                                & $105.92\pm15.31$                              & $<0.05$                                       & $<0.05$                                       & $>2000$        & $>2000$        \\
  060207         & $10.42\pm1.95$                                & $26.25\pm7.23$                                & $<0.08$                                       & $<0.15$                                       & $>300$         & $>200$         \\
  100424         & $0.26\pm0.04$                                 & $0.57\pm0.12$                                 & $<0.05$                                       & $<0.02$                                       & $>10$          & $>20$          \\
  151128         & $0.27\pm0.06$                                 & $2.44\pm0.99$                                 & $<0.08$                                       & $<0.14$                                       & $>30$          & $>20$          \\\hline
  050925         & $0.24\pm0.05$                                 & $1.16\pm0.37$                                 & $<0.06$                                       & $<0.18$                                       & $>20$          & $>10$           \\
  160220         & $0.42\pm0.06$                                 & $1.92\pm0.44$                                 & $<0.11$                                       & $<0.16$                                       & $>20$          & $>10$          \\
  140811         & $0.16\pm0.02$                                 & $0.57\pm0.13$                                 & $<0.05$                                       & $<0.06$                                       & $>10$          & $>10$          \\
  040610         & $0.12\pm0.01$                                 & $0.45\pm0.08$                                 & $<0.03$                                       & $<0.04$                                       & $>20$          & $>10$          \\
  \enddata

  \tablecomments{Observed fluxes in the 0.3--10~keV range computed
    using the time-integrated best-fit spectrum. The choice of model
    is based on the goodness of fit (Table~\ref{tab:stat}). The last
    two columns show the dynamic ranges of the transients.}
  
\end{deluxetable*}

The light curves in Figure~\ref{fig:res} show the temporal evolutions
of the transients. We note the significantly different durations and
timescales (Section~\ref{sec:sbo_mod}) ranging from
${\sim}$30--10,000~s. All transients show gradual rises that appear at
least marginally resolved, although quite steep in some cases. In
particular, no source shows a prompt ($<10$~s) rise followed by an
immediate decay. This has important implications for interpreting the
origins of the transients, since some are characterized by sharp rises
(Section~\ref{sec:alt}). We are not able to quantify the properties of
the decaying tails accurately since all transients quickly fade below
the detection limits. Dividing the light curves into different energy
bands is limited by the number of photons, and does not reveal any
significant behavior that is not captured by the time-resolved spectra
(Section~\ref{sec:res_spec}). Finally, we note in passing an apparent,
curious double-peak structure in the light curves of \xrt{}~070618 and
060207. We do not consider this further since it appears statistically
marginal, but a clearly double-peaked \xray{} transient has previously
been reported by \citet{jonker13}.

The average fluxes, peak fluxes, and constraints before and after the
transients are provided in Table~\ref{tab:lc} (with the peak fluxes
determined as described in Section~\ref{sec:obs}). This shows that the
transients have large dynamic ranges of at least 10 in all cases and
substantially higher in many cases. The constraints on the dynamic
ranges are primarily limited by the telescope sensitivity.

For two of the transients, we find some evidence for emission outside
the main transient event. This is the case for \xrt{}~161028, which is
marginally detected after the main event. This is likely the decaying
tail that extends beyond the selected time interval used to define the
duration. No sub-threshold source is present before the transient and
the limit indicates that any emission was lower than the marginal
detection after. We also find no indications of a source in two \xmm{}
observations\footnote{Observation IDs 0021140801 and 002114901.}
14~years before the transient.

\xrt{}~151219 shows a marginal quiescent flux level before the
transient. \xmm{} observed this position 7~months before and 18~months
after the transient\footnote{Observation IDs 0770380201 (before) and
  0802860201 (after).}. The last observation marginally detects a flux
level roughly half (consistent within uncertainties) of the level seen
just before the transient. We interpret this as quiescent \xray{}
emission from an underlying active galactic nucleus.
Using the flux measured just before the transient, the inferred
0.3--10~keV luminosity is $(1.4\pm{}0.4)\times{}10^{43}$\ergps{} for
$z=0.62$. No emission is detected in the earliest observation and just
after the transient in the main observation. This is likely due to the
limited sensitivity and inspections of the images reveal a
sub-threshold source present in both cases.

\subsection{Spectra}\label{sec:res_spec}
\begin{deluxetable*}{lccccccccccccc}
  \tablecaption{Results of Spectral Fits\label{tab:fit}}
  \tablewidth{0pt}
  \tablehead{
    \colhead{XT} & \colhead{$N_\mathrm{H,MW}$}     & \colhead{$N_\mathrm{H,BB}$}     & \colhead{$T$}          & \colhead{$T_1$}        & \colhead{$T_2$}        & \colhead{$N_\mathrm{H,PL}$}     & \colhead{$\Gamma$}     & \colhead{$\Gamma_1$}   & \colhead{$\Gamma_2$}   \\
    \colhead{}   & \colhead{($10^{20}$~cm$^{-2}$)} & \colhead{($10^{22}$~cm$^{-2}$)} & \colhead{(keV)}        & \colhead{(keV)}        & \colhead{(keV)}        & \colhead{($10^{22}$~cm$^{-2}$)} & \colhead{}             & \colhead{}             & \colhead{}} \startdata
  161028         & $1.9$                           & $<1.2$                          & $0.41_{-0.12}^{+0.17}$ & $0.53_{-0.12}^{+0.20}$ & $0.28_{-0.07}^{+0.11}$ & $<2.2$                          & $2.8_{-1.1}^{+1.8}$    & $2.5_{-0.7}^{+0.7}$    & $3.7_{-1.1}^{+1.3}$    \\
  151219         & $2.5$                           & $<1.8$                          & $0.37_{-0.11}^{+0.13}$ & $0.45_{-0.09}^{+0.15}$ & $0.30_{-0.05}^{+0.08}$ & $0.8_{-0.6}^{+1.0}$             & $3.5_{-1.2}^{+2.1}$    & $3.0_{-0.6}^{+0.7}$    & $4.3_{-0.8}^{+0.9}$    \\
  110621         & $3.0$                           & $1.3_{-1.2}^{+3.4}$             & $0.42_{-0.21}^{+0.27}$ & $0.52_{-0.13}^{+0.18}$ & $0.26_{-0.07}^{+0.11}$ & $3.4_{-2.0}^{+3.9}$             & $>2.9$                 & $4.1_{-1.1}^{+1.2}$    & $>5.5$                 \\
  030206         & $6.4$                           & $<0.4$                          & $0.47_{-0.05}^{+0.05}$ & $0.53_{-0.06}^{+0.06}$ & $0.38_{-0.05}^{+0.05}$ & $0.6_{-0.5}^{+0.5}$             & $3.4_{-0.5}^{+0.6}$    & $3.3_{-0.3}^{+0.3}$    & $3.9_{-0.5}^{+0.5}$    \\
  070618         & $1.6$                           & $<0.8$                          & $0.46_{-0.09}^{+0.10}$ & $0.64_{-0.10}^{+0.12}$ & $0.31_{-0.04}^{+0.05}$ & $0.8_{-0.3}^{+0.5}$             & $3.2_{-0.7}^{+0.8}$    & $2.5_{-0.3}^{+0.3}$    & $4.4_{-0.5}^{+0.6}$    \\
  060207         & $10.8$                          & $<1.6$                          & $0.93_{-0.21}^{+0.26}$ & $0.89_{-0.20}^{+0.33}$ & $0.62_{-0.14}^{+0.24}$ & $<3.9$                          & $1.9_{-0.8}^{+1.0}$    & $1.8_{-0.7}^{+0.7}$    & $2.4_{-0.7}^{+0.8}$    \\
  100424         & $5.1$                           & $<0.6$                          & $0.13_{-0.03}^{+0.03}$ & $0.13_{-0.03}^{+0.04}$ & $0.12_{-0.03}^{+0.03}$ & $<0.7$                          & $3.9_{-0.6}^{+2.8}$    & $4.0_{-0.9}^{+1.1}$    & $4.1_{-0.7}^{+0.9}$    \\
  151128         & $5.1$                           & $<4.2$                          & $0.16_{-0.11}^{+0.14}$ & $0.09_{-0.03}^{+0.06}$ & \nodata{}              & $<0.9$                          & $1.9_{-1.1}^{+3.2}$    & \nodata{}              & $1.4_{-1.4}^{+1.6}$    \\\hline
  050925         & $4.3$                           & $<1.9$                          & $0.45_{-0.17}^{+0.36}$ & $0.40_{-0.12}^{+0.23}$ & $0.59_{-0.37}^{+0.95}$ & $<2.4$                          & $2.1_{-0.8}^{+1.6}$    & $2.2_{-0.7}^{+0.8}$    & $1.8_{-1.2}^{+1.2}$    \\
  160220         & $8.4$                           & $<1.1$                          & $0.45_{-0.11}^{+0.13}$ & $0.40_{-0.08}^{+0.15}$ & $0.47_{-0.10}^{+0.14}$ & $0.5_{-0.4}^{+0.7}$             & $2.9_{-0.9}^{+1.4}$    & $3.0_{-0.6}^{+0.7}$    & $2.9_{-0.6}^{+0.7}$    \\
  140811         & $11.2$                          & $<4.7$                          & $0.32_{-0.13}^{+0.17}$ & $0.38_{-0.09}^{+0.16}$ & $0.28_{-0.05}^{+0.06}$ & $<3.7$                          & $2.7_{-0.9}^{+2.0}$    & $2.5_{-0.7}^{+0.9}$    & $3.0_{-0.6}^{+0.7}$    \\
  040610         & $4.2$                           & $<0.9$                          & $0.68_{-0.13}^{+0.17}$ & $0.96_{-0.20}^{+0.30}$ & $0.60_{-0.11}^{+0.14}$ & $0.4_{-0.4}^{+0.7}$             & $2.1_{-0.6}^{+0.7}$    & $1.8_{-0.5}^{+0.5}$    & $2.3_{-0.4}^{+0.4}$    \\
  \enddata

  \tablecomments{Parameters without integer subscripts are for the
    time-integrated spectra. The subscripts 1 and 2 refer to the first
    and second intervals of the time-resolved analysis,
    respectively. The Galactic absorptions ($N_\mathrm{H,MW}$) are
    frozen during the fits. The absorption column densities
    $N_\mathrm{H,BB}$ and $N_\mathrm{H,PL}$ are the redshifted host
    galaxy absorptions from the blackbody and power-law fits,
    respectively. Parameter estimates are missing for unconstrained
    fits.}
\end{deluxetable*}

The time-integrated spectral fits are shown in Figure~\ref{fig:res}
and the parameters from all fits are provided in Table~\ref{tab:fit},
including the time-resolved fits. Statistical measures of all fits are
provided in Table~\ref{tab:stat}. All fits are acceptable, although
there are indications that some spectral features are not properly
captured by the simple models. We do not attempt to fit more
complicated models due to the data quality and modeling
uncertainties. Additionally, the sources likely evolve in time, which
could introduce apparent spectral features in the time-integrated
spectra.

All transients are soft but show a range of different color
temperatures in the range 0.1--1~keV, or alternatively, photon indices
of 2--4. We note that there is an observational bias toward soft
sources due to the instrumental characteristics of \xmm{} and the
detection statistic, which scales with the number of photons rather
than fluence. The time-resolved fits show that the evolutions
generally go from harder to softer, although the significance is
marginal in some cases. No source shows any significant hardening.

Overall, the power-law model seems to result in slightly better
fits. It is possible that this is due to a stronger degeneracy between
the photon index and host absorption than between the temperature and
host absorption. This behavior can be seen from the fitted host
absorption column densities, which generally are lower for the
blackbody model than the power-law model. We do not attempt to
evaluate if the blackbody or the power-law model is more reliable in
this respect. This would require assuming an underlying SBO model,
modeling the absorption of galaxies, and computing the survey
sensitivity, which is beyond the scope of this paper. We simply
conclude that the photon indices might be unreliable and that the
degeneracy can result in apparently better fits when compared to the
blackbody model. Any uncertainties in the photon indices do not affect
our conclusions significantly since they are not used for inferring
physical parameters.

\section{SBO Interpretation}\label{sec:discussion}
In this section, we outline the SBO model (Section~\ref{sec:sbo_mod})
used for inferring physical parameters from the observables
(Section~\ref{sec:sn_prop}). We then discuss the effects of
asymmetries (Section~\ref{sec:asy}) and CSM (Section~\ref{sec:csm}),
which are not incorporated into the model. We conclude by commenting
on the individual SBO candidates and how the observations compare to
theory (Section~\ref{sec:not}).

\subsection{SBO Model}\label{sec:sbo_mod}
We use a model for SN SBOs to connect the observed properties to
physical parameters of the SN and its progenitor. The aim is to infer
the shock velocity ($v_\mathrm{sh}$), progenitor radius ($R$),
breakout density ($\rho$), and ejecta velocity ($v_\mathrm{ej}$). The
ejecta velocity is the typical bulk velocity defined as
\begin{equation}
  \label{eq:ej}
  v_\mathrm{ej}=\sqrt{\frac{E_\mathrm{exp}}{M_\mathrm{ej}}},
\end{equation}
where $E_\mathrm{exp}$ is the SN explosion energy and $M_\mathrm{ej}$
the ejecta mass. The observed parameters that constrain these
properties are the bolometric SBO energy ($E_\mathrm{SBO}$) and color
temperature ($T$).

Here, we closely follow the summary of SN SBO theory by
\citet{waxman17b}. For simplicity, we consider the non-relativistic
regime even though BSGs are expected to be in the transitional region
(${\sim}0.1c$) and possibly be mildly relativistic in some
cases. Non-relativistic SBOs are also more accurately modeled from a
theoretical perspective. More detailed SBO modeling is not motivated
due to the limited data and uncertainties in the redshifts.

The bolometric SBO energy is given by Equation~(31) of
\citet{waxman17b}
\begin{equation}
  \label{eq:ebo}
  E_\mathrm{SBO} = 2.2\times{}10^{47}R_{13}^{2}v_{\mathrm{sh},9}\kappa_{0.34}^{-1}~\text{erg},
\end{equation}
where $10^{n}Q_{n}~\mathrm{cgs}=Q$ in general for integer $n$ and
$0.34\kappa_{0.34}$~cm$^{2}$~g$^{-1} = \kappa$ is the opacity. The
opacity is given by $\kappa = (1+X)/5$~cm$^{2}$~g$^{-1}$, where $X$ is
the H mass fraction. This implies that $\kappa$ ranges from
0.2~cm$^{2}$~g$^{-1}$ (H\nobreakdash-poor) to 0.34~cm$^{2}$~g$^{-1}$
(H\nobreakdash-rich). The shock velocity is given by Equation~(27) of
\citet{waxman17b}
\begin{equation}
  \label{eq:vsbo}
  \frac{v_\mathrm{sh}}{v_\mathrm{ej}} = 13M_{\mathrm{ej},10}^{0.16}v_\mathrm{ej,8.5}^{0.16}R_{12}^{-0.32}\kappa_{0.34}^{0.16}f^{-0.05},
\end{equation}
where $3000v_\mathrm{ej,8.5}$~km~s$^{-1} = v_\mathrm{ej}$ and $f$ is a
numerical factor of order unity that depends on the detailed envelope
structure (Appendix~A of \citealt{calzavara04}). The numerical factor
for BSGs is given by Equation~(37) of \citet{sapir13}
\begin{equation}
  \label{eq:frho}
  \begin{split}
  f & = 0.072\mu_{0.62}^4L_{\star{},5}^{-1}M_{\mathrm{ej},10}^{3}\kappa_{0.34}^{-1}\\
    & \times{} \left(1.35-0.35L_{\star{},5}M_{\mathrm{ej},10}^{-1}\kappa_{0.34}\right)^4,
  \end{split}
\end{equation}
where $0.62\mu_{0.62}=\mu$ is the mean molecular weight and
$L_\star{}$ the progenitor luminosity. The parameter $\mu$ is given by
$\mu=(2X+0.75Y)^{-1}$, where $Y$ is the He mass fraction. Although the
time-integrated color temperature cannot be expressed analytically, it
can be fitted by Equation~(50) of \citet{waxman17b}
\begin{equation}
  \label{eq:tpeak}
  \begin{split}
  \log_{10}\left(\frac{3T}{\mathrm{eV}}\right) & = 1.4 + v_{\mathrm{sh},9}^{0.5} \\
  & + (0.25 - 0.05v_{\mathrm{sh},9}^{0.5})\log_{10}(\rho_{-9}),
  \end{split}
\end{equation}
where $\rho$ is given by Equation~(28) of \citet{waxman17b}
\begin{equation}
  \label{eq:rho}
  \rho =
  8\times{}10^{-9}M_{\mathrm{ej},10}^{0.13}v_\mathrm{ej,8.5}^{-0.87}R_{12}^{-1.26}
  \kappa_{0.34}^{-0.87}f^{0.29}~\text{g~cm}^{-3}.
\end{equation}

The observed data are not able to constrain all parameters, in
particular those that the observables are relatively insensitive
to. Therefore, we take some values based on \sna{} because SBOs should
be most easily detectable from BSGs similar to \sna{} \citep{sapir13,
  sapir14, waxman17b}. Specifically, we assume
$M_\mathrm{ej}=15$~\Msun{}\citep{utrobin19, menon19, alp19},
$L_\star{}=150,000$~\Lsun{} \citep{woosley87}, and a
H\nobreakdash-rich surface composition of $X=0.7$ and $Y=0.3$. To
illustrate the weak dependence on these parameters, we note that
choosing a H\nobreakdash-poor star with a factor of 10 lower
$M_\mathrm{ej}$ and a factor of 10 higher $L_\star{}$ only changes $R$
by a factor of 0.83, $v_\mathrm{sh}$ by 0.86, $\rho$ by 4.0, and
$v_\mathrm{ej}$ by 1.7. With the fiducial \sna{} values,
Equations~\eqref{eq:ebo}, \eqref{eq:vsbo}, \eqref{eq:tpeak}, and
\eqref{eq:rho} form a system of equations with 4 unknowns. The
solution process can be simplified by solving Equation~\eqref{eq:vsbo}
for $v_\mathrm{ej}$, inserting it into Equation~\eqref{eq:rho}, and
using Equation~\eqref{eq:ebo} to express $\rho$ in terms of
$E_\mathrm{SBO}$, which results in
\begin{equation}
  \label{eq:rho2}
  \rho = 5\times{}10^{-9}E_{\mathrm{SBO},46}^{-0.75}~\text{g~cm}^{-3}
\end{equation}
for our assumed parameter values.

In addition to the constraint on $R$ from the model above, an
independent constraint can be placed based on the light curve. The
light curve shape is initially determined by the light travel time
from different parts of the progenitor. The light curve rises and
remains approximately constant until a time of $t_{R/c} \equiv R/c$,
after which it starts decaying \citep{nakar10}. This light travel time
argument only applies if the SBO is at least approximately
spherical. Furthermore, it is only valid if $t_{R/c}$ is longer than
the dynamical time of the breakout shell (breakout shell width over
$v_\mathrm{sh}$). This is expected to be the case for all but the most
extended RSGs, which are likely to be too cool to be observed.

\subsection{Inferred SN Properties}\label{sec:sn_prop}
\begin{deluxetable*}{lccccccccccc}
  \caption{Inferred SBO Properties\label{tab:inferred}}
  \tablewidth{0pt}
  \tablehead{
    \colhead{XT} & \colhead{$t_{R/c}$} & \colhead{$L_\mathrm{peak}$}        & \colhead{$E_\mathrm{SBO}$} & \colhead{$R_{t}$} & \colhead{$R_{E}$} &   \colhead{$v_\mathrm{sh}$} &   \colhead{$v_\mathrm{ej}$} &                 \colhead{$\rho$} & \colhead{$E_\mathrm{exp}$} \\
    \colhead{}   & \colhead{(s)} & \colhead{($10^{44}$~erg~s$^{-1}$)} & \colhead{($10^{46}$~erg)}  & \colhead{(\Rsun)} & \colhead{(\Rsun)} & \colhead{($10^{3}$\kmps{})} & \colhead{($10^{3}$\kmps{})} & \colhead{($10^{-9}$~g~cm$^{3}$)} & \colhead{($10^{51}$~erg)}} \startdata
  161028         &            42 & $3.39_{-1.03}^{+2.40}$             & $0.3_{-0.1}^{+0.2}$        &                14 &                11 &                          22 &                         1.5 &                             13.0 &                        0.7 \\
  151219         &            90 & $32.88_{-11.00}^{+53.76}$          & $2.4_{-0.8}^{+4.0}$        &                24 &                30 &                          25 &                         2.2 &                              2.5 &                        1.4 \\
  110621         &           380 & $0.39_{-0.22}^{+7.45}$             & $0.5_{-0.3}^{+9.5}$        &               150 &                14 &                          24 &                         1.7 &                              8.2 &                        0.9 \\
  030206         &           360 & $172.47_{-23.77}^{+28.72}$         & $79.0_{-10.9}^{+13.2}$     &                71 &               146 &                          35 &                         4.5 &                              0.2 &                        6.0 \\
  070618         &            70 & $48.78_{-10.68}^{+17.05}$          & $7.6_{-1.7}^{+2.6}$        &                22 &                49 &                          30 &                         2.9 &                              1.1 &                        2.6 \\
  060207         &           110 & $6.44_{-2.03}^{+2.74}$             & $2.2_{-0.7}^{+0.9}$        &                36 &                23 &                          39 &                         3.0 &                              2.8 &                        2.6 \\
  100424         &          5600 & $0.05_{-0.01}^{+0.12}$             & $1.0_{-0.3}^{+2.6}$        &              2136 &                30 &                          11 &                         1.1 &                              4.9 &                        0.3 \\
  151128         &          1800 & $1.83_{-1.08}^{+56.98}$            & $2.5_{-1.5}^{+78.1}$       &               524 &                41 &                          14 &                         1.5 &                              2.5 &                        0.6 \\\hline
  050925         &           850 & $0.21_{-0.08}^{+0.14}$             & $0.3_{-0.1}^{+0.2}$        &               282 &                10 &                          24 &                         1.6 &                             12.9 &                        0.7 \\
  160220         &           750 & $0.51_{-0.12}^{+0.23}$             & $0.6_{-0.2}^{+0.3}$        &               249 &                16 &                          25 &                         1.8 &                              6.9 &                        1.0 \\
  140811         &          1800 & $0.93_{-0.41}^{+4.64}$             & $2.9_{-1.3}^{+14.6}$       &               494 &                34 &                          23 &                         2.1 &                              2.2 &                        1.3 \\
  040610         &          3200 & $0.32_{-0.07}^{+0.09}$             & $1.8_{-0.4}^{+0.5}$        &               919 &                23 &                          34 &                         2.6 &                              3.1 &                        2.0 \\
  \enddata

  \tablecomments{Timescales ($t_{R/c}$) are defined by hand
    (Section~\ref{sec:sbo_mod}) and are given in the observer
    frame. The redshift correction is applied when computing the
    radius $R_t=ct/(1+z)$. The luminosities and energies are
    bolometric. The parameters $L_\mathrm{peak}$ and $E_\mathrm{SBO}$
    are observational and do not rely on the SBO modeling. The
    intervals for $L_\mathrm{peak}$ and $E_\mathrm{SBO}$ only
    represent the \xray{} fitting uncertainties and do not include the
    redshift uncertainties. The large uncertainties for \xrt{}~151128
    are due to a combination of few photons and a low lower
    temperature limit. We refrain from estimating the uncertainties
    for the remaining parameters since they include uncertainties in
    the redshifts as well as the SBO modeling, which are difficult to
    quantify (Sections~\ref{sec:hosts} and~\ref{sec:sbo_mod}).}
\end{deluxetable*}

Here, we use the model outlined above to infer physical SN parameters,
which are provided in Table~\ref{tab:inferred}. The table includes
values that are derived using both the energetics and the light curve
shape. The primary goal is to determine if the obtained parameters are
consistent with typical SN SBO values. This is mainly relying on
comparisons with theoretical predictions due to the lack of
observations. We do not attempt to model and propagate all
uncertainties into these parameters. However, we stress that both the
redshift (Section~\ref{sec:hosts}) and modeling uncertainties are
large. More detailed analyses of the transients require spectroscopic
redshifts and will be the subject of future studies.

From the light curves (Figure~\ref{fig:res}), we define the typical
timescales by hand, as described above. We denote radii determined
using this method $R_t$ to distinguish $R_t$ from the radii inferred
from the energetics, which we denote $R_E$.

The peak luminosities ($L_\mathrm{peak}$) and total SBO energies
($E_\mathrm{SBO}$) rely on the redshift estimates and extrapolations
of the spectra. We use the time-integrated best-fit blackbody models
to extrapolate from the observed 0.3--10~keV range to obtain the
bolometric quantities. For the typical best-fit temperatures of
${\sim}$0.4~keV, the 0.3--10~keV range already captures the vast
majority of the bolometric flux. The time-averaged luminosities
obtained from the fits are scaled to the light curve peaks using the
peak-to-average flux ratio (Section~\ref{sec:obs};
Table~\ref{tab:lc}). Having computed the energy $E_\mathrm{SBO}$, it
is straightforward to solve for the remaining parameters
(Section~\ref{sec:sbo_mod}). Finally, we note that $L_\mathrm{peak}$
is more uncertain than $E_\mathrm{SBO}$, both from an observational
and theoretical perspective. Therefore, we focus on $E_\mathrm{SBO}$
for the current analysis and provide $L_\mathrm{peak}$ solely for
comparisons with other \xray{} transients in general.

The parameter values in Table~\ref{tab:inferred} are in reasonable
agreement with predictions for SN SBOs. For reference, we compute the
expected values for a SBO similar to \sna{}. In addition to the
assumed values in Section~\ref{sec:sbo_mod}, we complete the model by
taking $R=40$~\Rsun{} and $E_\mathrm{exp}=1.5\times{}10^{51}$~erg
\citep{mccray93}. For this set of parameters, the derived properties
are: $E_\mathrm{SBO} = 4.0\times10^{46}$~erg,
$v_\mathrm{sh}=23,000$\kmps{}, $v_\mathrm{ej}=1800$\kmps{}, and
$\rho=1.7\times{}10^{-9}$~g~cm$^{-3}$.

The values for $E_\mathrm{SBO}$, $R_E$, $\rho$, and $E_\mathrm{exp}$
are generally within an order of magnitude of the predicted
values. These are dependent and primarily show that the observed
$E_\mathrm{SBO}$, and to a lesser extent $T$, are within the range of
predictions for SBOs. Estimates of $v_\mathrm{sh}$ and
$v_\mathrm{ej}$ are more closely clustered around $20,000$\kmps{} and
$2000$\kmps{}, respectively. This is expected because the velocities
are mainly constrained by, and quite insensitive to, $T$ through
Equations~\eqref{eq:vsbo} and~\eqref{eq:tpeak}. This implies that most
soft spectra would result in reasonable $v_\mathrm{sh}$ and
$v_\mathrm{ej}$. Furthermore, there is a strong observational bias
against detecting objects cooler than 0.1~keV because of the lower
0.3~keV energy limit of \xmm{} and ISM absorption.

It is worth pointing out that the independent estimates of $R$ are
within a factor of ${\sim}$2 for 5 of the sources. These two estimates
are completely independent and lend some strength to the SBO
interpretation. The remaining objects all have substantially larger
$R_t$ than $R_E$. This is a consequence of the observed timescale
being longer than the expected timescale. Possible reasons for longer
timescales are asymmetries and dense CSM structures, as discussed
below.

\subsection{Asymmetries}\label{sec:asy}
Strong asymmetries in core-collapse SNe are predicted from theory
\citep{janka16, muller16, couch17} and clearly supported by
observations \citep{larsson16, abellan17, grefenstette17}. Asymmetric
SBOs have been studied \citep{suzuki10, couch11}, but have relied on
bimodal, jet-like, two-dimensional axisymmetric explosions. These
asymmetries were introduced by hand but showed that asymmetries could
contribute to longer timescales and significantly affect the light
curve shape. To first order, the asymmetries only affect when the
shock reaches the surface. The minor differences in shock velocity and
breakout angle should have much smaller effects on the SBO
properties. This means that only the light curve shape should be
affected by the asymmetries.

More recently, \citet{wongwathanarat15} performed self-consistent
three-dimensional neutrino-driven SN simulations to late times past
SBO (Figure~14 of \citealt{wongwathanarat15}). They used a set of RSG
and BSG single-star progenitors and have now also studied BSGs that
are the results of binary mergers (\citealt{menon17};
A.~Wongwathanarat 2020, private communication). Although these
simulations do not compute the SBO emission, they track the shock as
it propagates through the star and breaks the surface. For the
single-star BSG progenitor B15, the time difference between the first
and last parts of the shock breaking the surface is approximately
200~s. This is a relatively spherical model. The BSG merger
progenitors are more asymmetric and show time differences of
${\sim}$1000~s. These values can be compared to the times required for
the shock to propagate through the stars of 4500--7500~s. For W15,
which is a relatively asymmetric RSG, the fastest shock breaks out at
${\sim}$70,000~s and the slowest at ${\sim}$83,000~s post-bounce.

We emphasize that these asymmetries are not introduced by hand and
develop spontaneously during the explosion from initial seed
perturbations. This implies that asymmetries are expected to develop
even for spherical progenitors. It is also worth pointing out that
aspherical SBOs could both shorten or lengthen the observed timescale,
depending on the viewing angle.

To summarize, it is likely that asymmetric shocks develop during the
core-collapse process. These asymmetries significantly affect the
duration and light curve shape of the observed SBO. We tentatively
conclude that asymmetries can introduce time variations of up to
$20$\,\% of the shock crossing time of the star, which is an order of
magnitude longer than the light crossing time $R/c$ in extreme cases.
The effects of asymmetries on the timescale are sufficient to reduce
the tensions between $R_t$ and $R_E$ in the cases where they are
significantly different.

\subsection{Circumstellar Medium}\label{sec:csm}
The CSM affects the SBO emission if the optical depth of the CSM is
larger than $c/v_\mathrm{sh}$. In these cases, the shock does not
break out at the surface, but instead propagates into the CSM. A
notable example is the initial \xray{} transient associated with
SN~2008D \citep{soderberg08}, which is particularly interesting
because of similarities with some of our SBO candidates. For
reference, the SN~2008D SBO had a peak luminosity of
$4\times{}10^{43}$\ergps{}, total energy of $6\times{}10^{45}$~erg, a
timescale of $t_{R/c}\approx{}150$~s, and a $\Gamma\approx 2$
power-law spectrum \citep{modjaz09}. A thermal spectrum with a
temperature of 0.75~keV fits the data worse but is still statistically
acceptable. From these values and using our methods, we would infer
$R_t=65$~\Rsun{} and $R_E=11$~\Rsun{}.

A common interpretation of the SN~2008D \xray{}
transient is a SBO from the explosion of a WR star surrounded by a
thick wind \citep{chevalier08, balberg11, svirski14}. CSM SBOs are
also often used to explain observations of Type~IIn and super-luminous
SNe (Section~6.1 of \citealt{waxman17b} and references
therein). However, these SBOs are very different and are not expected
to appear as \xray{} transients on timescales that our searches are
sensitive to.

We do not attempt to explore the effects of a dense CSM in detail due
to the wide range of possible scenarios and the observational
uncertainties. Instead, we simply note that a dense CSM could increase
the timescale and that some of our candidates are similar to
SN~2008D.

\subsection{Notes on Individual Objects}\label{sec:not}
Even though the inferred parameters (Table~\ref{tab:inferred}) agree
reasonably well with theoretical predictions, there are some aspects
that cause tensions with a clean SBO interpretation. In light of the
discussions of asymmetries and effects of the CSM, we provide more
detailed comments on individual objects below.
\begin{itemize}
\item The first two objects, \xrt{}~161028 and 151219, agree very well
  with BSG SBO predictions. The derived quantities are close to
  fiducial SBO values under the assumptions of spherical symmetry and
  thin CSM. The only parameter that stands out is $E_\mathrm{SBO}$ for
  \xrt{}~161028, which is an order of magnitude lower than the
  reference value. Given the uncertainties, we do not consider this to
  be a major problem and conclude that both objects are strong SN SBO
  candidates.
\item All properties of \xrt{}~110621\footnote{After submission of our
    paper, \citet{novara20} reported an independent discovery of
    \xrt{}~110621 (EXMM 023135.0$-$603743 using their
    notation). Importantly, they find a spectroscopic redshift of
    0.092 for the host galaxy (compared to our value of 0.095), which
    eliminates much of the uncertainty for our analysis of
    \xrt{}~110621. Overall, their and our results and interpretations
    agree very well with each other. There are minor differences in
    the spectral fits, which likely are due to a combination of
    different methods, the high background (Section~\ref{sec:obs}),
    and the high host absorption.} agree well with predictions for BSG
  SBOs, except for the timescale. This can be seen from the
  discrepancy between $R_t=150$ and $R_E=14$. One possible explanation
  for the longer observed timescale is an asymmetric BSG SBO. Another
  possibility is a SBO from a WR with a dense wind, similar to the
  SN~2008D SBO. They both have practically the same peak luminosities
  and total energies, although the light curve shapes are slightly
  different and \xrt{}~110621 has a softer spectrum. Given the
  expected variances in light curves and spectra, a similar origin for
  \xrt{}~110621 is possible. We are not able to distinguish between an
  asymmetric BSG SBO and a dense CSM WR SBO, but conclude that both
  are possible origins.
\item \xrt{}~030206 is the most energetic object and would require
  extreme SN parameters. Interestingly, the inferred $E_\mathrm{exp}$
  of $6\times{}10^{51}$~erg is in the range of
  $(7.1\pm5.4)\times{}10^{51}$~erg typical of Type~Ic\nobreakdash-BL
  SNe \citep{taddia19}. Our modeling does not strictly apply to these
  SNe and any firm conclusions would require further analysis
  (Section~\ref{sec:grbs}).
\item Both \xrt{}~070618 and 060207 agree well with the BSG SBO
  interpretation. The main uncertainty for these objects is due to the
  lack of host detections. This implies that the redshift estimates
  are highly uncertain, which propagates into all inferred parameters.
\item \xrt{}~100424 and 151128 are the coolest objects and are much
  slower than expected for typical SBOs from BSGs. The observed
  properties suggest that they might instead be associated with SBOs
  from RSGs. A less explored possibility is SBOs from yellow
  supergiants (\citealt{smartt09, ergon14}), which in the current
  context effectively is an intermediate class between RSGs and
  BSGs. The interpretations for these two objects are mainly driven by
  the timescales and temperatures, which are observationally
  reliable. The total energies are more than an order of magnitude
  lower than predicted for RSG SBOs, but are uncertain due to the
  observed energy range relative to the temperatures, as well as the
  redshift uncertainties. The transients are also hotter than RSG SBO
  predictions, but this is anticipated because of the very strong
  observational biases against the cooler majority of RSG SBOs. The
  timescales are also slightly longer than expected, but this can be
  alleviated by invoking asymmetries.
\item The last four transients, \xrt{}~050925, 160220, 140811, and
  040610 have longer timescales than expected for BSG SBOs. They are
  all largely similar to \xrt{}~110621 (and implicitly also the
  SN~2008D SBO) and the same arguments apply to these four
  sources. The main differences for these last four transients are the
  uncertain host redshifts and the risk of being foreground
  contamination (Section~\ref{sec:dwarfs}).
\end{itemize}

\section{Alternative Explanations}\label{sec:alt}
In this section, we explore possible sources other than SNe that could
produce the observed transients. Section~3 of \citet{bauer17} provides
a similar discussion with a slightly different focus. They discuss
CDF\nobreakdash-S~XT1, which suffers from similar potential
contaminants as our SBO candidates. The most likely sources to be
confused for SN SBOs are flares from Galactic late-type dwarfs
(Section~\ref{sec:dwarfs}). We discuss GRBs as potential sources in
Section~\ref{sec:grbs} and a number of less probable sources in
Section~\ref{sec:other}. We note that the predicted number of SN SBOs
\citep{sapir14, waxman17b} is comparable to the number we observe,
which lends strength to the SBO interpretation.

\subsection{Dwarf Stars}\label{sec:dwarfs}
\xray{} flares from late-type dwarf stars \citep{gudel04, gudel09b,
  benz10} could potentially be confused for SBO candidates, in cases
where a host galaxy is not clearly detected. The coolest object
observed to flare in \xray{}s is an L1~dwarf \citep{de_luca20}. A
qualitative difference between SN SBOs and dwarf flares is that dwarf
flares are recurrent (and any dwarfs with multiple flares would
already have been excluded from our sample). Faint flares are frequent
and occur up to tens of times per day, while the rate of the most
powerful flares are limited by the observed sample size
\citep{loyd18b, loyd18}.

The distribution of peak \xray{} (0.3--10~keV) luminosities of dwarf
flares extends to ${\sim}10^{30}$\ergps{} and decreases steeply with
peak flux \citep{cook14, williams14, loyd18b, loyd18, de_luca20}. The
effective temperatures for the majority of the flares are in the range
${\sim}$0.5--2~keV \citep{pallavicini90, robrade05, pye15, de_luca20}.
Typical timescales range from a few hundred to a few thousand seconds,
and the dynamic ranges vary from a factor of 2 to more than 300
\citep{favata00, pandey08, robrade10, pye15}.

For our purposes, we focus on the extreme cases despite their low
rates. The 2014 April 23 flare from DG~CVn is the most luminous
observed M~dwarf flare \citep{caballero-garcia15, osten16}. DG~CVn is
a binary system composed of two M4Ve dwarfs with a combined absolute
\jband{} magnitude of 7.2\magab{}. The flare reached a peak flux of
$3\times{}10^{32}$\ergps{} and the dominant flare component evolved on
a timescale of ${\sim}300$~s. Another comparable event is the 2008
April 25 EV~Lac flare \citep{osten10}. EV~Lac is an M3.5V~dwarf with
an absolute \jband{} magnitude of 8.5\magab{}. The \xray{} flare
reached a peak flux of $10^{32}$\ergps{} and the dominant component
evolved on a timescale of ${\sim}$300~s. However, these events are
very hard, with peak temperatures of 25~keV (DG~CVn) and 10~keV
(EV~Lac), and even triggered the gamma-ray burst monitor \swift{}/BAT
\citep{barthelmy05}. The L1~dwarf flare reported by \citet{de_luca20}
reached a peak flux of $6.3\times{}10^{29}$\ergps{}, had a temperature
of 1.4~keV, and evolved on timescales of ${\sim}$3~ks. The absolute
\jband{} magnitude of the L1~dwarf is 12.8\magab{}. Finally, we note
that \citet{glennie15} report a tentative association of an \xray{}
transient with an L1~dwarf. The inferred peak luminosity is
$7\times{}10^{31}$\ergps{}; however, both the spectral type and
distance to the source are uncertain.

We can use the optical and NIR data (Table~\ref{tab:host_mag}) to
constrain the distance to a potential dwarf star. This can be combined
with the peak \xray{} fluxes (Table~\ref{tab:lc}) to infer a peak
\xray{} luminosity if the source is a dwarf. For reference, the
\xray{} luminosity expressed in terms of peak flux
($F_{-13}=10^{13}F_\mathrm{peak}$\ergpspcm{}), apparent magnitude
($m$), and absolute magnitude ($M$) is
\begin{equation}
  \label{eq:lim}
  L_\mathrm{peak} =  1.2F_\mathrm{-13}\times{}10^{27+0.4\times{}(m - M)}\text{\ergps{}}.
\end{equation}
The peak flare flux relative to absolute magnitude is comparable for M
and L~dwarfs. For the discussion, we adopt fiducial dwarf absolute
magnitudes of $M_r=9$\magab{} \citep{bochanski11, pecaut13} and
$M_J=8$\magab{} \citep{pecaut13, carnero_rosell19}, which are
representative of an M4~dwarf, and similarly for an L1 dwarf:
$M_z=14.3$\magab{} \citep{carnero_rosell19, de_luca20} and
$M_J=13$\magab{}.

For two of the three candidates without hosts, \xrt{}070618 and
\xrt{}060207, the inferred luminosities are orders of magnitude higher
than what have been observed from dwarf stars. For the last candidate
without a host, \xrt{}040610, the observations are marginally
consistent with an L1~dwarf because of the weak NIR constraints. The
strongest constraint is in the \zband{}, which implies a peak
luminosity of $6\times{}10^{29}$\ergps{} for a distance of
330~pc. This is similar to the L1~flare reported by \citet{de_luca20}.

We now turn our attention to the three last SBO candidates with
possible host associations: \xrt{}~050925, 160220, and 140811. These
have been classified as galaxies based on morphology
(Table~\ref{tab:host_par}) but are potentially blends of two point
sources, implying that the transient would be a flare from one of the
foreground stars. By comparing the optical and NIR photometry with
\xray{} data, we find that all three are consistent with being dwarf
flares for M~dwarfs at ${\sim}3$~kpc or L~dwarfs at
${\sim}300$~pc. However, this would require them to be among the most
luminous dwarf flares for either spectral type. Furthermore, if these
sources indeed consist of two blended sources, then using the combined
magnitude of the two sources as we have done above would underestimate
the true \xray{} flux.

Dust could, in principle, obscure the emission of an optical
source. However, this is challenging to reconcile with the H column
density through the Milky Way (Table~\ref{tab:fit},
\citealt{willingale13}). Furthermore, the observed \xray{} spectra
generally indicate very low levels of optical and NIR absorption
\citep{predehl95}.

To summarize, it is possible that at least some of the bottom four
candidates: \xrt{}~050925, 160220, 140811, and 040610, are dwarf
flares. However, we are not able to firmly conclude this without
additional data.

\subsection{Gamma-Ray Bursts}\label{sec:grbs}
The prompt GRB phase evolves on timescales of 0.3~s for short GRBs and
10~s for long GRBs \citep{meegan96, von_kienlin20}. The spectra
typically peak at energies in the range 100--1000~keV
\citep{goldstein12}, and extend into soft X-ray energies
\citep{villasenor05, lien16}. Both types are followed by afterglows
with luminosities in the range $10^{47}$--$10^{51}$\ergps{} at 100~s
that decay on timescales of hours to days \citep{berger14,
  bauer17}. The spectral shape in the 0.3--10~keV range is a power law
with photon indices in the range ${\sim}$1.5--3 \citep{zhang07,
  racusin09b}.

If the SBO candidates are related to GRBs, the peak luminosities have
to be much lower than typical GRB afterglows (Figure~4 of
\citealt{bauer17}). For the candidates with host galaxies, it is
highly unlikely that the true redshifts are substantially higher than
estimated. The lack of a sharply rising prompt phase, overall light
curve shapes, and spectra are also difficult to reconcile with
standard on-axis GRBs.

GRBs could potentially agree with some of our observations if observed
far off-axis. This applies to both short \citep{sun19, xue19, dado20}
and long GRBs \citep{dado19}. The effects of a large viewing angle are
longer duration, much lower peak luminosity, lower spectral peak, and
lower fluence. The models for off-axis GRBs can be adjusted to agree
qualitatively with our observations; however, many model parameters
are essentially unconstrained and can be tuned to produce a very large
range of observed timescales, spectra, and energies.

Another source of emission is the hot cocoon, which is the result of
interactions between the jet and surrounding material. This is
produced in both short \citep{lazzati17} and long \citep{suzuki13,
  de_colle18} GRBs. The predictions for both types are \xray{}
transients with peak luminosities around $10^{46}$--$10^{48}$\ergps{}
and fast rise times, with the details being quite sensitive to the
off-axis angle. Cocoon emission from short GRBs is expected to evolve
on timescales of $<30$~s with temperatures of 0.2--2~keV. The
corresponding values for long GRBs are ${\sim}$100~s and
${\sim}$0.2~keV. We note that these values agree reasonably well with
the most luminous of the observed transients, \xrt{}~030206,
particularly for a long GRB cocoon.

\subsection{Other Possibilities}\label{sec:other}
There are other astrophysical sources that could give rise to \xray{}
transients. Many of these are expected to also produce persistent
optical emission. Therefore, we note that the optical constraints
discussed in Section~\ref{sec:dwarfs} (Equation~\ref{eq:lim}) are
general and apply to any type of object, such as a companion star in
an \xray{} binary. It is also important to consider the Galactic
latitude of the SBO candidates (Table~\ref{tab:identifiers}). Galactic
objects are generally confined to the Galactic plane, except for
objects that are expected to be very close, such as late-type
dwarfs. Furthermore, the prior probability of a neutron star to align
with an extragalactic source is $<1$\,\% for a neutron star solid
angle density of 1000~deg$^{-2}$ \citep{ofek09, sartore10}. Here, we
very briefly mention a number of potential sources without exploring
all details. We simply note that the following classes are able to
produce some, but not all, of the observables without significant
fine-tuning or requiring very rare circumstances.

Tidal disruption events \citep{rees88, phinney89} produce soft \xray{}
transients that typically evolve on timescales of 10--300~days
\citep{komossa15b, kochanek16}. This scenario involves an ordinary
star being accreted onto a supermassive black hole, which also
determines the timescale. This is much longer than the SBO candidates,
but disruptions of white dwarfs by intermediate mass black holes have
been suggested to produce tidal disruption events on timescales of
${\sim}$10~ks \citep{fernandez19, shen19, maguire20}.

\xray{} binaries \citep{nooraee13, walter15, martinez-nunez17} can
produce a number of different types of \xray{}
flares. Type\nobreakdash-I (thermonuclear) \xray{} bursts
\citep{galloway17, galloway20} have peak luminosities of
${\sim}3\times10^{38}$\ergps{} and evolve on timescales of 10--300~s,
but can extend up to several thousand seconds in some cases. Ordinary
accretion processes and type\nobreakdash-II outbursts
\citep{van_den_eijnden17} generally do not produce the high dynamic
ranges that characterize flares. So-called supergiant fast \xray{}
transients \citep{romano15, ducci19, sguera20} are ${\sim}3$~ks flares
with dynamic ranges of ${\sim}10^{4}$ and peak luminosities on the
order of $10^{37}$\ergps{}, which are driven by wind accretion from OB
supergiant companions. Although \xray{} binaries can produce a range
of \xray{} transients, they are necessarily associated with bright
companions that are challenging to reconcile with our optical and NIR
data.

White dwarf binaries \citep{mukai17} produce variable \xray{}
emission through accretion and surface nuclear burning
\citep{schwarz11, morii13, starrfield16, ness19}. They can produce
very soft spectra and evolve on short timescales, but typically do not
flare with sufficiently large dynamic ranges to be confused with SN
SBOs.

Magnetars \citep{turolla15, mereghetti15, kaspi17} are isolated
neutron stars that are characterized by very strong magnetic
fields. They are known to produce high-energy transient emission, most
notably so-called giant flares \citep{hurley05, palmer05}. These are
much harder than the SBO candidates and have light curves with a more
prompt burst of ${\sim}1$~s followed by a much fainter decaying tail,
similar to short GRBs.

\section{Summary and Conclusions}\label{sec:conclusions}
When a star undergoes core collapse, a shock is launched from the
core. This shock propagates through the star and when it reaches the
surface, a burst of soft \xray{}s is released. This is the first
electromagnetic signal that escapes a core-collapse SN and it carries
independent information about the progenitor radius, asymmetries, and
final mass-loss history. The \xray{} transients are characterized by
total energies on the order of $10^{45}$--$10^{47}$~erg, timescales of
10--1000~s, and soft spectra corresponding to temperatures of
$0.03$--$3$~keV. These values depend sensitively on the type of
progenitor. RSGs generally produce slower and cooler SBOs, BSG SBOs
are intermediate, and SBOs from WR progenitors are faster and hotter.

We search the \xmm{} archive for serendipitously observed SN
SBOs. This archive is likely to contain more SN SBOs than any other
because of the large effective area, large FoV, and 20 year lifetime
of \xmm{}. Our search results in 12 SN SBO candidates. They are all
consistent with being SBOs, but it is possible that a few are Galactic
foreground dwarf stars or produced by other types of extragalactic
sources. We focus on the SN SBO interpretation and investigate the
inferred physical properties. In addition to the \xray{} data, we
analyze public wide-field optical and NIR data and find host galaxy
candidates for nine of the sources. We estimate host redshifts of
0.1--0.6, except for one source at redshift 1.17.

The SBO candidates have energies on the order of $10^{46}$~erg,
timescales ranging from minutes to hours, and soft spectra with color
temperatures of 0.1--1~keV. Using these observables, we are able to
infer progenitor and SN properties. Two candidates are probably SBOs
from BSGs. A third candidate is similar but slightly slower, which
could be interpreted as an asymmetric BSG SBO or breakout from a wind
surrounding a more compact WR progenitor (similar to SN~2008D). There
is one transient with a total energy of $8\times{}10^{47}$~erg, which
is higher than SBO predictions. This could potentially be the SBO from
an extreme SN. There are two more sources that are likely BSG SBOs,
but lack host galaxy identifications and are more uncertain. Two
candidates appear to be extreme cases of RSG or possibly yellow
supergiant SBOs. Finally, the last four candidates could either be
asymmetric BSG SBOs, WR wind SBOs, or Galactic foreground
contamination.

Many of the SBO candidates show signs of significant asymmetries or
optically thick CSM. This is consistent with both theory and other
types of observations of core-collapse SNe. Asymmetries arise
spontaneously during the core-collapse process and primarily affect
the timescale of the SBO. We emphasize that this very naturally
explains a broader diversity in SBO durations (up to 20\,\% of the
shock crossing time) than expected from spherical models. This
additional variance agrees well with our observations. The effect of
CSM on the SBO candidates is more difficult to disentangle due to the
very large parameter space, which can give rise to a wide range of
observables. More detailed conclusions need further modeling of the
individual sources, deeper optical data, and spectroscopic
redshifts. Optical follow-up data of the SNe would also ideally be
obtained. This will be the subject of future studies.

Future \xray{} observations will detect an ever-increasing number of
SN SBOs. The \xray{} instrument currently most likely to detect SBOs
is \erosita{} \citep{predehl10, predehl17}. We predict \erosita{} to
detect 2 SN SBOs per year using Equations~43 of \citet{sapir13} and~97
of \citet{waxman17b}. This is a back-of-the-envelope estimate where we
also attempt to correct for ISM absorption and the diversity in
observed properties. We conclude by emphasizing the importance of live
analyses of future \xray{} data and rapid follow-up observations to
fully capitalize on these rare opportunities.

\acknowledgments{We are grateful to Annop Wongwathanarat for providing
  measures of SBO asymmetries from SN simulations, and Eleni Tsaprazi
  for large-scale structure density contrasts. We also thank Claes
  Fransson and Jens Jasche for helpful discussions, and the anonymous
  referee for the comments. This work was supported by the Knut and
  Alice Wallenberg Foundation. This research has made use of data
  obtained from the 3XMM \xmm{} serendipitous source catalogue
  compiled by the 10 institutes of the \xmm{} Survey Science Centre
  selected by ESA. This research has made use of data produced by the
  EXTraS project, funded by the European Union's Seventh Framework
  Programme under grant agreement no 607452. This research has made
  use of data obtained through the High Energy Astrophysics Science
  Archive Research Center Online Service, provided by the NASA/Goddard
  Space Flight Center. This research has made use of the SIMBAD
  database, operated at CDS, Strasbourg, France. This research has
  made use of the VizieR catalogue access tool, CDS, Strasbourg,
  France. This research has made use of ``Aladin sky atlas'' developed
  at CDS, Strasbourg Observatory, France. This research made use of
  \texttt{hips2fits}\footnote{https://alasky.u-strasbg.fr/hips-image-services/hips2fits},
  a service provided by CDS. This research has made use of NASA's
  Astrophysics Data System.}

\vspace{5mm}
\facilities{\textit{XMM} (EPIC)}

\software{
  ADS \citep{kurtz00},
  Aladin \citep{bonnarel00, boch14},
  astropy (3.0.4; \citealt{astropy13, astropy18}),
  FTOOLS \citep{blackburn95},
  HEAsoft (6.26.1; \citealt{heasarc14}),
  Le PHARE (2.2; \citealt{arnouts99, ilbert06}),
  matplotlib (2.0.2; \citealt{hunter07}),
  numpy (1.13.1; \citealt{van_der_walt11}),
  SAOImage DS9 (8.1; \citealt{joye03}),
  SAS (18.0.0; \citealt{gabriel04}),
  scipy (1.1.0; \citealt{virtanen20}).
  SIMBAD \citep{wenger00},
  VizieR \citep{ochsenbein00},
  XSPEC (12.10.1f; \citealt{arnaud96}).
}

\appendix

\section{Searching the \xmm{} Observations}\label{app:search}
We use two different algorithms to search for SN SBOs in archival
\xmm{} data. The aim of both search algorithms is solely to identify
\xray{} transients. Rejecting spurious detections and careful source
identification is performed in a subsequent stage
(Section~\ref{sec:classification}). For these reasons, we do not
attempt to correct for instrumental effects such as vignetting,
deadtime, or chip gaps in the transient detection process. Some
details related to data preparation are also not considered, such as
extraction regions falling outside of CCDs and nearby sources located
within background regions. These effects are managed in the proper
data reduction (Section~\ref{sec:obs}) for the final SBO candidates.

Notably, we do not exclude periods of high background when searching
for transients because transients can occur during these times and
could, in principle, be securely identified. The high background is
produced by protons in the Earth's magnetosphere and depends on the
satellite altitude and level of solar activity \citep{carter07}. These
proton flares affect 30 to 40\,\% of the total \xmm{} observation
time. Excluding these time periods increases the S/N of typical
persistent sources, but this is not true for a transient source that
is only detectable during a background flare. The challenge is to
distinguish an interesting astrophysical transient from background
flares. This distinction is difficult to do solely using light curves,
but can easily be done in images because background flares affect the
CCDs uniformly whereas a transient source appears as a point.

\subsection{Custom Transient Source Finder}\label{app:search_custom}
Our custom algorithm starts from the Processing Pipeline Subsystem
(PPS) event lists and finds all variable sources. We use all 11,500
public observations with EPIC imaging archived at HEASARC as of 2019
November 11. Since we are expecting faint and soft sources, we combine
the data from all EPIC cameras \citep{struder01, turner01} and
restrict the energy range to 0.3--2~keV.

The first step of the search algorithm is to bin the event list along
the spatial and temporal dimensions. Effectively, this creates a
series of images (a data cube) for each observation, which is divided
into three-dimensional ``cells'' with two spatial and one temporal
dimension. The spatial binning is $20\times{}20$~arcsec$^2$, which is
chosen to contain the cores of the PSFs of EPIC. The encircled energy
at a radius of 10~arcsec is around 60\,\% for EPIC. The PSF wings are
much broader and can be neglected. The temporal binning is performed
at 5 different timescales ranging from 100~s to 10,000~s with
logarithmic spacing.  To mitigate the effects of the discretizations
in space and time, we repeat the binning by shifting the grid by half
the bin size along each dimension. The total number of combinations
introduced by the shifts and temporal binning at different timescales
is 40 ($2\times{}2\times{}2\times{}5$). This means that 40 data cubes
are created for each observation.

The second step is to find transients in the data cubes. To do this,
we perform a simplified statistical test for variability in each cell
relative to the grid cells before and after. Before and after in this
context refers to the cells at the same spatial position, but one bin
earlier and later along the temporal dimension,
respectively. Importantly, the transient detection algorithm must be
able to handle the highly temporally varying background and reject
slowly varying sources. For each cell, we first need to estimate the
expected persistent source flux at the position. This is done by
linearly interpolating the background-subtracted fluxes of the cells
before and after in time. The background subtraction in the before and
after cells are performed by subtracting the average of the four
spatially neighboring grid cells at the respective times. To prevent
over-subtraction (negative counts), we simply force the expected
source flux in the current bin to be at least 1 count. This provides
the expected persistent source flux at the current position. To this
flux, we add the average background of the four spatially neighboring
pixels. This sum is the expected total count rate in the current
cell. Finally, we assume the count rate to be Poissonian and compute
the $p$-value to obtain the observed count rate given the expected
rate. This $p$-value should capture any variability in excess of
linear variability in time, while also being insensitive to the highly
temporally varying background.

To separate the variable sources, we perform a cut based on the
variability measure. As a test statistic, we rescale the $p$-value and
use the negative log-likelihood $-\ln(p)$. We select a limiting
statistic of 25, which leaves a very low probability of noise passing
the filter. This cut leaves 380 detections, which are investigated by
eye in subsequent steps.

\subsection{Finding Transients in \cat{}}\label{app:search_cat}
The \xmm{} Survey Science Centre routinely publishes catalogs of
detected sources, the latest being \cat{} at the time
of writing. This catalog contains observations from 2000 February 3
through 2017 November 30 and contains 775,153 detections of 531,454
unique sources.
The first step toward identifying SBOs is analyzing the source light
curves. \cat{} includes light curves, but those are not suited for
fast transients because they were created with a common bin-width with
a minimum number of photons in every bin \citep[Section~5.2
of][]{watson09}. Consequently, the time interval before the rise of a
transient source (and possibly also after the transient has faded)
results in an increased common bin-width. This increase could
drastically reduce the sensitivity for short, bright flare-like
transients. For these reasons, we decide to construct custom light
curves for our SBO search.\footnote{We note that the Exploring the
  \xray{} Transient and variable Sky (EXTraS)
  project \citep{de_luca16} aim to provide more detailed temporal data
  products for all \xmm{} sources. However, the EXTraS data for
  aperiodic short-term variability (EXTraS Working Package~2) only
  includes data that were publicly available by 2012 December 31
  (corresponding to 3XMM\nobreakdash-DR4). The EXTraS light curves are
  also intended for general variable behavior and are not optimized
  for fast and possibly faint transients. For these reasons, we only
  briefly inspected the EXTraS WP2 catalog, and we find no objects
  that were not already detected by our other methods.}

We use the positions from the \cat{} catalog of identified sources and
generate light curves for all 550,614 detections that are more than
15\degr{} outside of the Galactic plane. We do not search the Galactic
plane because we found no candidates in this region using our custom
finder and because robust source identification is very difficult in
dense fields. The increased \xray{} absorption close to the Galactic
plane also drastically reduces the possibility of detecting
SBOs. Furthermore, we discard very faint sources by omitting all
objects with a detection likelihood reported by \cat{} as less than
10. This likelihood is based on the likelihood ratio described by
\citet{cash79}. The likelihood is defined as $-\ln(P)$, where $P$ is
the probability of the detection occurring by chance \citep[formally,
the probability of the null hypothesis; Section~4.4.3 of][]{watson09}.

To create the light curves, we download the \xmm{} PPS event lists and
combine all exposures of the EPIC pn, MOS1, and MOS2 cameras within
each observation and limit the energy range to 0.3--10~keV. From the
combined event list, we extract source events from a circular region
with a radius of 20~arcsec centered on the source position (taken from
\cat{}). Background events are then extracted from an annular region
centered on the source with an inner radius 50~arcsec and an outer
radius of 100~arcsec. The light curves are binned such that each bin
contains 25 counts from the source region. This binning requirement
ignores the background count rate and results in bins with different
temporal widths. The reason for choosing this dynamic-binning approach
is to be able to effectively capture the large variations in flux for
transient sources. The background events are binned to the same
temporal bins as the source light curves and are then subtracted from
the source to create a final light curve.

The next step is to find all transient sources. We identify sources
showing transient-like behavior by requiring that they fulfill at
least one of the following heuristics:
\begin{itemize}
\item The ratio of the maximum background-subtracted flux bin over the
  50th flux percentile (i.e.\ percentile of the bins weighted by time
  for this individual light curve) is larger than 3, while the
  signal-to-background ratio (S/B) is higher than 10 at the time of
  peak flux.
\item Same as above, but with a peak flux a factor of 5 above the 50th
  percentile and a S/B of at least 3.
\item At least 10~ks of the background-subtracted light curve is
  within 1$\sigma$ of 0\ctsps{}. Moreover, at least one bin has a
  source flux higher than 0.05\ctsps{} with a S/B higher than 3.
\end{itemize}
The combination of all heuristics is constructed to distinguish
flare-like sources while being insensitive to background flares and
inaccurate background subtraction. All heuristics are then also
re-evaluated with the modification that the time intervals when the
background count rate is higher than 0.05\ctsps{} are
discarded.\footnote{We note that the limit of 0.05\ctsps{} refers to
  the rate of our custom background light curves. These light curves
  are different from the background light curves in the standard
  \xmm{} data reduction, which are customarily defined as the total
  rate in the entire FoV within 10--12~keV for the pn CCD and above
  10~keV for the MOS CCDs.} Approximately 11,000 sources pass these
criteria and are investigated by eye at a subsequent stage. This can
be compared to \cat{}, which classifies 5934 sources as variable. The
difference in the number of variable sources is not surprising because
of our simplified background treatment, which results in a very large
number of sources being classified as variable due to inaccurately
subtracted background flares. Furthermore, these numbers are not
strictly comparable because we apply a number of filters and have
defined variability differently.

\section{Host Photometry}\label{app:host_pho}
Table~\ref{tab:host_mag} provides the optical and NIR data for the
candidate host galaxies. The multi-band optical data are used for the
redshift SED fitting. Specifically, we only use optical data from a
single survey for each source to avoid systematic uncertainties
between different surveys. It is not possible to perform reliable SED
fits using only NIR data, but the NIR data are still useful since they
provide an important check for the inferred absolute magnitudes.

\section{Fit Statistics}\label{app:fit}
Table~\ref{tab:stat} provides statistic measures for all spectral
fits. We note that the number of photons (essentially the degrees of
freedom) is very limited in some cases, especially for the
time-resolved spectra. This introduces substantial variance in the
goodness measures.

\begin{longrotatetable}
  \begin{deluxetable}{lcccccccccccccc}
    \tablecaption{Host Photometry\label{tab:host_mag}}
    \tablewidth{0pt}               
    \tablenum{B.1}
    \tablehead{\colhead{\xrt{}} & \colhead{$m_u$} &  \colhead{$m_g$} &  \colhead{$m_r$} &  \colhead{$m_i$} &  \colhead{$m_z$} &                   \colhead{$m_Y$} &  \colhead{$m_J$} & \colhead{$m_H$} & \colhead{$m_K$} & \colhead{Source} \\
                     \colhead{} & \colhead{(mag$_\mathrm{AB}$)} &  \colhead{(mag$_\mathrm{AB}$)} &  \colhead{(mag$_\mathrm{AB}$)} &  \colhead{(mag$_\mathrm{AB}$)} &  \colhead{(mag$_\mathrm{AB}$)} &                   \colhead{(mag$_\mathrm{AB}$)} &  \colhead{(mag$_\mathrm{AB}$)} & \colhead{(mag$_\mathrm{AB}$)} & \colhead{(mag$_\mathrm{AB}$)} &       \colhead{} }\startdata
    161028                      &  $22.2\pm{}0.3$ & $21.46\pm{}0.32$ & $20.69\pm{}0.06$ & $20.50\pm{}0.10$ & $20.40\pm{}0.29$ &                         \nodata{} &         $>18.01$ &        $>17.79$ &        $>17.15$ &     SDSS16/2MASS \\
    151219                      &  $23.1\pm{}0.2$ & $22.37\pm{}0.05$ & $21.49\pm{}0.02$ & $20.88\pm{}0.06$ & $20.94\pm{}0.20$ &                  $20.68\pm{}0.30$ & $20.57\pm{}0.35$ &  $20.3\pm{}0.3$ &  $19.6\pm{}0.2$ &      KiDS/VIKING \\
    110621                      &       \nodata{} & $19.68\pm{}0.01$ & $19.63\pm{}0.01$ & $19.34\pm{}0.01$ & $19.39\pm{}0.02$ &                  $19.38\pm{}0.07$ & $19.17\pm{}0.13$ &       \nodata{} &  $19.9\pm{}0.3$ &          DES/VHS \\
    030206                      &       \nodata{} & $21.89\pm{}0.11$ & $22.05\pm{}0.12$ & $22.37\pm{}0.11$ & $21.73\pm{}0.15$ &                           $>21.3$ &         $>18.01$ &        $>17.79$ &        $>17.15$ & Pan-STARRS/2MASS \\
    070618                      &       \nodata{} &          $>24.3$ &         $>24.08$ &         $>23.44$ &         $>22.69$ &                          $>21.44$ &         $>18.01$ &        $>17.79$ &        $>17.15$ &        DES/2MASS \\
    060207                      &         $>17.9$ &          $>18.0$ &          $>18.0$ &          $>18.0$ &          $>18.0$ &                         \nodata{} &         $>21.11$ &       \nodata{} &        $>19.95$ &    SkyMapper/VHS \\
    100424                      &       \nodata{} &  $21.4\pm{}0.06$ & $20.37\pm{}0.04$ & $19.97\pm{}0.02$ & $19.73\pm{}0.03$ &                  $19.33\pm{}0.07$ &         $>18.01$ &        $>17.79$ &        $>17.15$ & Pan-STARRS/2MASS \\
    151128                      &       \nodata{} &  $21.5\pm{}0.10$ & $20.30\pm{}0.04$ & $19.83\pm{}0.03$ & $19.47\pm{}0.03$ &                  $19.63\pm{}0.09$ &         $>18.01$ &        $>17.79$ &        $>17.15$ & Pan-STARRS/2MASS \\\hline
    050925                      &         $>17.9$ &          $>18.0$ &          $>18.0$ &          $>18.0$ &          $>18.0$ &                  $21.53\pm{}0.57$ & $20.69\pm{}0.51$ &       \nodata{} &        $>20.35$ &    SkyMapper/VHS \\
    160220                      &         $>17.9$ &          $>18.0$ &          $>18.0$ &          $>18.0$ &          $>18.0$ &                         \nodata{} & $20.25\pm{}0.28$ &        $>17.79$ &        $>17.15$ &    SkyMapper/VHS \\
    140811                      &       \nodata{} & $21.80\pm{}0.06$ & $20.61\pm{}0.02$ & $19.71\pm{}0.01$ & $19.32\pm{}0.02$ &                  $19.02\pm{}0.04$ &         $>18.01$ &        $>17.79$ &        $>17.15$ & Pan-STARRS/2MASS \\
    040610                      &       \nodata{} &          $>22.9$ &          $>22.8$ &          $>22.7$ &          $>21.9$ &                           $>20.9$ &         $>18.01$ &        $>17.79$ &        $>17.15$ & Pan-STARRS/2MASS \\
    \enddata
  
    \tablecomments{All magnitudes are converted \citep{hewett06,
        blanton07} to the AB magnitude system \citep{oke83}. All
      detections use either \citet{petrosian76} or \citet{kron80}
      magnitudes, which are suitable for extended sources. We note
      that the upper limits are given for different thresholds:
      3$\sigma$ (2MASS), 5$\sigma$ (Pan-STARRS and VHS), S/N of 10
      (DES), and 10$\sigma$ (SkyMapper). The limits are for point
      sources except for the DES limits, which are for apertures of
      1.95~arcsec. For reference, \cite{metcalfe13} find that galaxies
      have a 50\,\% completeness about 0.4\magab{} brighter than stars
      using Pan-STARRS observations. Filters of different telescopes
      are also slightly different.}

    \tablerefs{
      Sloan Digital Sky Survey Data Release 16 (SDSS-IV DR16, \citealt{york00, ahumada20}),
      Two Micron All Sky Survey (2MASS, \citealt{skrutskie06}),
      Kilo-Degree Survey Data Release 3 (KiDS-ESO-DR3, \citealt{de_jong13, de_jong17}),
      Visible and Infrared Survey Telescope for Astronomy (VISTA) Kilo-degree Infrared Galaxy Survey Data Release 2 (VIKING DR2, \citealt{edge13}),
      Dark Energy Survey Data Release 1 (DES DR1, \citealt{abbott18}),
      VISTA Hemisphere Survey Data Release 4.1 (VHS DR4.1, \citealt{mcmahon13}),
      Panoramic Survey Telescope and Rapid Response System Data Release 2 (Pan-STARRS DR2, \citealt{chambers16}),
      SkyMapper \citep{wolf18}.}

  \end{deluxetable}
\end{longrotatetable}

\begin{longrotatetable}
  \begin{deluxetable}{lccccccccccccc}
    \tablecaption{Fitting Statistics\label{tab:stat}}
    \tablewidth{0pt}
    \tablenum{C.1}
    \tablehead{
      \colhead{\xrt{}} & \colhead{$C_\mathrm{BB}/$DoF} & \colhead{$G_\mathrm{BB}$} & \colhead{$C_\mathrm{BB,1}/$DoF} & \colhead{$G_\mathrm{BB,1}$} & \colhead{$C_\mathrm{BB,2}/$DoF} & \colhead{$G_\mathrm{BB,2}$} & \colhead{$C_\mathrm{PL}/$DoF} & \colhead{$G_\mathrm{PL}$} & \colhead{$C_\mathrm{PL,1}/$DoF} & \colhead{$G_\mathrm{PL,1}$} & \colhead{$C_\mathrm{PL,2}/$DoF} & \colhead{$G_\mathrm{PL,2}$}} \startdata
    161028             & $44/43=1.02$                  &                      0.82 & $18/20=0.92$                    &                        0.59 & $29/19=1.55$                    &                        0.98 & $41/43=0.95$                  &                      0.66 & $17/20=0.86$                    &                        0.47 & $30/19=1.55$                    &                        0.98 \\
    151219             & $71/64=1.11$                  &                      0.94 & $43/31=1.39$                    &                        0.97 & $35/32=1.10$                    &                        0.83 & $66/64=1.03$                  &                      0.81 & $41/31=1.31$                    &                        0.89 & $33/32=1.04$                    &                        0.71 \\
    110621             & $55/53=1.04$                  &                      0.61 & $17/24=0.70$                    &                        0.09 & $32/25=1.29$                    &                        0.78 & $56/53=1.05$                  &                      0.62 & $17/24=0.72$                    &                        0.10 & $32/25=1.29$                    &                        0.79 \\
    030206             & $187/209=0.89$                &                      0.96 & $123/133=0.93$                  &                        0.95 & $103/93=1.11$                   &                        0.99 & $169/209=0.81$                &                      0.58 & $118/133=0.88$                  &                        0.87 & $99/93=1.07$                    &                        0.96 \\
    070618             & $141/177=0.80$                &                      0.88 & $84/106=0.79$                   &                        0.81 & $61/74=0.82$                    &                        0.76 & $128/177=0.72$                &                      0.45 & $77/106=0.73$                   &                        0.43 & $60/74=0.81$                    &                        0.69 \\
    060207             & $42/48=0.88$                  &                      0.47 & $23/24=0.97$                    &                        0.54 & $11/20=0.57$                    &                        0.18 & $39/48=0.81$                  &                      0.31 & $23/24=0.96$                    &                        0.48 & $10/20=0.51$                    &                        0.09 \\
    100424             & $80/80=1.01$                  &                      0.69 & $36/41=0.89$                    &                        0.39 & $42/42=1.00$                    &                        0.52 & $75/80=0.94$                  &                      0.33 & $36/41=0.89$                    &                        0.26 & $36/42=0.86$                    &                        0.19 \\
    151128             & $51/34=1.50$                  &                      1.00 & $29/19=1.52$                    &                        0.98 & $17/11=1.55$                    &                        0.96 & $45/34=1.32$                  &                      0.90 & $28/19=1.49$                    &                        0.78 & $15/11=1.39$                    &                        0.87 \\\hline
    050925             & $40/32=1.24$                  &                      0.81 & $28/20=1.41$                    &                        0.90 & $10/9=1.12$                     &                        0.48 & $35/32=1.08$                  &                      0.50 & $24/20=1.19$                    &                        0.64 & $8/9=0.94$                      &                        0.33 \\
    160220             & $82/83=0.99$                  &                      0.56 & $43/40=1.09$                    &                        0.69 & $30/41=0.73$                    &                        0.09 & $76/83=0.92$                  &                      0.27 & $39/40=0.96$                    &                        0.34 & $30/41=0.74$                    &                        0.07 \\
    140811             & $108/88=1.23$                 &                      0.99 & $24/24=1.00$                    &                        0.67 & $64/62=1.03$                    &                        0.82 & $101/88=1.15$                 &                      0.83 & $22/24=0.93$                    &                        0.39 & $63/62=1.02$                    &                        0.56 \\
    040610             & $117/141=0.83$                &                      0.30 & $26/37=0.70$                    &                        0.15 & $101/102=0.99$                  &                        0.66 & $110/141=0.78$                &                      0.07 & $18/37=0.50$                    &                        0.02 & $102/102=1.00$                  &                        0.55 \\
    \enddata

    \tablecomments{Quantities denoted by $C$ are fit statistics and
      $G$ are goodness measures (Section~\ref{sec:tech}). The
      subscripts are BB for blackbody and PL for power law, and first
      (1) and second (2) interval for time-resolved
      spectra. Quantities without subscript integers are
      time-integrated values. DoF refers to the degrees of freedom.}
  \end{deluxetable}
\end{longrotatetable}

\bibliography{main}{}
\bibliographystyle{aasjournal}

\end{document}